\documentclass{emulateapj}

\shorttitle{RADIO JET IN NGC 7479}
\shortauthors{LAINE AND BECK}
\begin{document}

\submitted{ACCEPTED FOR PUBLICATION IN THE ASTROPHYSICAL JOURNAL}
\title{RADIO CONTINUUM JET IN NGC 7479}

\author{Seppo Laine}
\affil{{\it Spitzer} Science Center - Caltech, MS 220-6,
Pasadena, CA 91125}
\email{seppo@ipac.caltech.edu}

\and

\author{Rainer Beck}
\affil{Max-Planck-Institut f\"{u}r Radioastronomie, Auf dem
H\"{u}gel 69, 53121 Bonn, Germany} \email{rbeck@mpifr-bonn.mpg.de}

\begin{abstract}

The barred galaxy NGC~7479 hosts a remarkable jet-like radio
continuum feature: bright, 12-kpc long in projection, and hosting an
aligned magnetic field. The degree of polarization is 6\%--8\% along
the jet, and remarkably constant, which is consistent with helical
field models. The radio brightness of the jet suggests strong
interaction with the ISM and hence a location near the disk plane.
We observed NGC~7479 at four wavelengths with the VLA and Effelsberg
radio telescopes. The equipartition strength is 35--40~$\mu$G for
the total and $\ge10~\mu$G for the ordered magnetic field in the
jet. The jet acts as a bright, polarized background. Faraday
rotation between 3.5 and 6~cm and depolarization between 6 and 22~cm
can be explained by magneto-ionic gas in front of the jet, with
thermal electron densities of $\simeq$~0.06~cm$^{-3}$ in the bar and
$\simeq$~0.03~cm$^{-3}$ outside the bar. The regular magnetic field
along the bar points toward the nucleus on both sides. The regular
field in the disk reveals multiple reversals, probably consisting of
field loops stretched by a shearing gas flow in the bar. The
projection of the jet bending in the sky plane is in the sense {\em
opposite\/} to that of the underlying stellar and gaseous spiral
structure. The bending in 3-D is most easily explained as a
precessing jet, with an age less than 10$^{6}$ years. Our
observations are consistent with very recent triggering, possibly by
a minor merger. NGC~7479 provides a unique opportunity to study
interaction-triggered 15-kpc scale radio jets within a spiral
galaxy.

\end{abstract}


\section{INTRODUCTION}
\label{s:intro}

Radio jets are commonly seen around radio galaxies, which themselves
are usually associated with distant elliptical galaxies
\citep{brid84}. Relatively few examples of radio galaxies are known
at distances less than 100~Mpc from us. The radio jets in these
systems can have lengths up to a few Mpc. The radio continuum
synchrotron emission comes from relativistic electrons spiraling in
magnetic fields. The origin of the power in radio galaxies is tied
to energetic events taking place near their nuclei, most likely
ejections or outflows from an accretion disk surrounding a massive
nuclear black hole \citep[e.g.,][]{rees82}.

Smaller-scale jets are often seen near the nuclei of Seyfert galaxies
\citep[e.g.,][and references therein]{ho01}. In these objects the jets
have lengths from less than a parsec to a few kiloparsecs. Again, the
energy source is probably near the nuclear black hole, this time
10--1000 times less massive than the black holes in radio galaxies.
For example, NGC~3079 hosts a nuclear radio jet feature within the
central parsec \citep{irwin88} and bipolar radio lobes extending out
to 3~kpc, perpendicular to the edge-on galaxy plane \citep{duric88}.
The Circinus spiral galaxy shows similar kpc-scale (projected) bipolar
radio lobes \citep{harnett90}, and a projected 100-pc scale jet-like
feature aligned with the radio lobes \citep{elmo98}, together with a
parsec-scale nuclear outflow \citep{green03}. The spiral galaxy
0313-192 hosts a one-sided nuclear jet of about 2~kpc length which
powers bright radio lobes of type FR~I extending to about 100~kpc from
the nucleus \citep{ledlow01,keel06}.

Curiously, there exist relatively few nearby galaxies with large jets
on the 10-kpc scale. Disk galaxies with such jets are even less
common. In fact, the only examples of 15~kpc-scale anomalous radio
arms or double-sided radio jets in disk galaxies in the nearby
Universe are in NGC~4258 \citep*[e.g.,][]{hummel89,krause04} and
NGC~7479 \citep{laine98}. The radio jet in NGC~4258 is detected out to
about 14 kpc radius and is of spiral form, trailing with respect to
the galaxy rotation. It is clearly interacting with the dense ISM gas
in the galactic plane, evidenced by similar structures seen in
H$\alpha$ and X-ray emission \citep*[e.g.,][]{wilson01} and a peculiar
distribution of the CO line emission in the disk
\citep*{krause90,krause07}. Numerous investigations have been carried
out on the nature of the jet in NGC~4258
(e.g.,\citeauthor{cecil95} \citeyear{cecil95}; \citeauthor{cox96} \citeyear{cox96}; \citeauthor{herrn97} \citeyear{herrn97}; \citeauthor{vogler99} \citeyear{vogler99}; \citeauthor{cecil00} \citeyear{cecil00}; \citeauthor{wilson01} \citeyear{wilson01}).
\citet{wilson01} argue that the interaction could be indirect via halo
gas heated by the jet and falling back onto the disk, so that the jet
itself may be extraplanar. In contrast to the jet in NGC~4258, the
bright, jet-like radio continuum structure in NGC~7479 has remained
practically unnoticed since its serendipitous discovery as a byproduct
of the \ion{H}{1} observations of \citet{laine98}.

NGC~7479 is a relatively nearby (distance $\approx 32$~Mpc) SBc type
spiral galaxy. Table~\ref{ngc7479tab} gives some basic properties of
NGC~7479. The nucleus has been classified as a LINER \citep{keel83} and
as a Seyfert 1.9 \citep*{ho97}. The Chandra X-ray observations by
\citet{panessa06} revealed an X-ray luminosity of
1.3$\times$10$^{41}$~ergs~cm$^{-2}$~s$^{-1}$ in the 2--10~keV band. The
inferred nuclear black hole mass is 1.2$\times$10$^{7}$~M$_{\sun}$,
which is almost as large as the nuclear black hole in NGC~4258 (4$\times$10$^{7}$~M$_{\sun}$;
\citeauthor{miyoshi95} \citeyear{miyoshi95}; \citeauthor{greenhill95a}
\citeyear{greenhill95a}; \citeauthor{greenhill95b}
\citeyear{greenhill95b}; \citeauthor{herrn99} \citeyear{herrn99}). The
X-ray  luminosity of NGC~4258 in the 2--10~keV band is
7.2$\times$10$^{40}$ ergs~cm$^{-2}$~s$^{-1}$ \citep{panessa06}, very
close to the value measured for NGC~7479. Therefore, the $L_{\rm
X}$/$M_{\rm BH}$ ratio is about a factor of 6 higher in NGC 7479
compared to NGC~4258. The VLA observations of NGC~7479 at 2~cm with
$\simeq $~0\farcs15 resolution \citep*{nagar05} show a
$2.5$~mJy~beam$^{-1}$ nuclear source in NGC~7479. VLA observations at
6~cm and 20~cm with $\simeq $~1\arcsec\ resolution reveal an extended
feature pointing from the nucleus towards the north-west
\citep{ho01,laine06}, coinciding with the inner part of the radio jet
(see below). Optical {\it HST} images (G. F. Benedict, unpublished)
show a complicated dust structure with a stronger dust lane leading
into the nucleus from north (apparently an inner extension of the
leading bar dust lane).

At larger scales, there are many signs of a recent perturbation in
the disk of NGC~7479. These include the strongly asymmetric spiral
structure, consisting of a strong and long western spiral arm that
appears to bifurcate west of the nucleus, while there is little
organized spiral structure on the eastern side
\citep[e.g.,][]{quillen95}. Further signatures of a recent
perturbation include the irregular \ion{H}{1} velocity field along
the western arm \citep{laine98}, a peculiar perturbation in the
H$\alpha$ velocity field north of the nucleus \citep{laine99},
anomalous dust lane structure around the bar as seen in the optical
images \citep{laine96}, and emission line excitation asymmetries
\citep*{martin00}. These characteristics have led \citet{quillen95}
and \citet{laine99} to suggest that the galaxy has recently suffered
a minor merger with a sizable companion. No companions have been
found with optical imaging \citep{saraiva03} or \ion{H}{1}
observations within a radius of several tens of kiloparsecs around
NGC~7479 \citep{laine98}. There is no visible trace of a remnant
of a merging companion either, but the minor merger models of
\citet{laine99} predict that the nuclear remnant of a mostly
disrupted companion should currently be in the bar, north of the
nucleus, coinciding with the perturbation in the H$\alpha$ velocity
field. The strong stellar bar was also proposed to have formed as a
result of the minor merger.

The first radio continuum polarization observations of NGC~7479 by
\citet{beck02} with the VLA at several wavelengths with 30~arcsec
resolution could not resolve the jet, but showed a curious
wing-like pattern at 6~cm, extending perpendicular to the bar,
with magnetic field lines oriented parallel to the bar.

This paper reports observations of polarized radio continuum
emission from NGC~7479, and specifically on the anomalous jet-like
radio continuum structure that has a total projected extent of
around 12~kpc and forms a spiral that {\it leads} with respect to
the galaxy rotation. In other words, the radio continuum feature
bends in the opposite sense to the stellar and gaseous spiral arms.
Since its discovery by \citet{laine98}, no evidence of this feature
at any other spectral regime has been found. This suggests that the
feature is not interacting with the gas in the disk of NGC~7479, and
likely extends (at least partly) out of the plane, while its
high radio brightness indicates strong interaction with the ISM in
the disk. We present total intensity and polarized emission data
that further suggest a jet origin for the anomalous radio continuum
feature. The origin, orientation, and nature of the jet are examined,
and compared to radio jets seen in other nearby galaxies. We also
investigate the nuclear point source, the bar region, the extended
radio continuum emission from NGC~7479, and the nature of point
sources within and outside the optical disk with the help of
polarization measurements.

\begin{table*}
\center
\caption{BASIC PARAMETERS OF NGC 7479.\label{ngc7479tab}}
\begin{tabular}{lc}
\tableline
\tableline
Parameter & Value \\
\tableline Galaxy type & SBbc(s) \\
Environment & Isolated \\
Right Ascension (J2000.0) & 23$^h$ 04$^m$ 56\fs 65 \\
Declination (J2000.0) & 12\degr 19\arcmin 22\farcs 4 \\
Galactic Longitude (J2000.0) & 86.2711\degr \\
Galactic Latitude (J2000.0) & -42.8417\degr \\
Heliocentric velocity & 2371 km~s$^{-1}$ \\
Distance & 32 Mpc \\
Linear scale & 1 arcsec $\cong$ 160 pc \\
Optical size at B$_{25}$ & 4.1 arcmin x 3.1 arcmin \\
Inclination & 51\degr \\
Position angle of the disk major axis & 22\degr \\
Integrated \ion{H}{1} mass & 8.6 $\times$~10$^{9}$~M$_{\sun}$\\
H$_{2}$ mass & 2.5~$\times$~10$^{10}$~M$_{\sun}$ \\
Central activity & LINER; Sy~1.9; starburst \\
$M_{\rm bh}$ & 1.2$\times$10$^{7}$~$M_{\sun}$ \\
Deprojected bar length (radius) & 8.6 kpc \\
Position angle of bar & 6\degr \\
\tableline
\end{tabular}

\tablecomments{Galaxy type from \citeauthor{sand87}
\citeyear{sand87}, Right Ascension and Declination from NED,
heliocentric velocity from \citeauthor{laine98} \citeyear{laine98},
distance assuming Hubble law with {\it H}$_{\rm 0}$~=
75~km~s$^{-1}$, size from NED, inclination, position angle and
\ion{H}{1} mass from \citeauthor{laine98} \citeyear{laine98},
H$_{2}$ mass from \citeauthor{young89} \citeyear{young89}, LINER
classification from \citeauthor{keel83} \citeyear{keel83}, Sy1.9
classification from \citeauthor{ho97} \citeyear{ho97}, starburst
classification from \citeauthor{dev89} \citeyear{dev89}, black hole
mass from \citeauthor{panessa06} \citeyear{panessa06}, bar length
from \citeauthor{blackman83} \citeyear{blackman83}, position angle
of the bar from \citeauthor{burbidge60} \citeyear{burbidge60}.}
\end{table*}

\section{OBSERVATIONS}
\label{s:sampleobs}

Very Large Array (VLA) observations of NGC~7479 were taken during
several separate observing sessions (Table~\ref{vlaobs}). Weather
conditions were generally good during the observations. Absolute
flux calibration is based on scans of 3C~286 and 3C~138, which also
served as absolute polarization angle calibrators. The target galaxy
observations were bracketed between 1--2 minute scans of the nearby
phase calibrator J2253+1608, and their fluxes were bootstrapped to
those of the primary calibrators. Uncertainties in the fluxes quoted
in the VLA Calibrator Manual are dominated by the uncertainty in
setting the absolute flux scale, and are estimated to be $\approx
5\%$. Uncertainties in the positions are estimated to be less than
0\farcs 1. Uncertainties in the absolute value of the polarization
angle are estimated to be less than 1\degr.

\begin{table*}
\center
\caption{PARAMETERS OF THE OBSERVING RUNS. \label{vlaobs}}
\begin{tabular}{lccccccc}
\tableline
Date & Telescope & Obs. time (sec) & Wavelength (cm) & Notes \\
\tableline
\tableline
1996 Mar 22 & VLA-C & 2400 & 18, 22 & Beck et al. (2002)\\
1996 Jul 22 & VLA-D & 2200 & 3.5 & Beck et al. (2002) \\
1996 Jul 22 & VLA-D & 1800 & 6.2 & Beck et al. (2002)\\
1997 Jul 2  & VLA-C & 5300 & 18, 22 & Beck et al. (2002)\\
1998 Jan 9--10 & VLA-D & 19500 & 3.5 & \\
1999 Apr 9  & VLA-D & 8500 & 6.2 & Hardware failure\\
2001 Aug 2  & VLA-C & 25500 & 3.5 & \\
2001 Aug 24 & VLA-C & 24790 & 6.2 & \\
2003 Feb 21, 23 & Effelsberg & 12000 & 3.6 & \\
\tableline
\end{tabular}
\end{table*}

\begin{figure*}[th]
\centering
\includegraphics[width=6.5in]{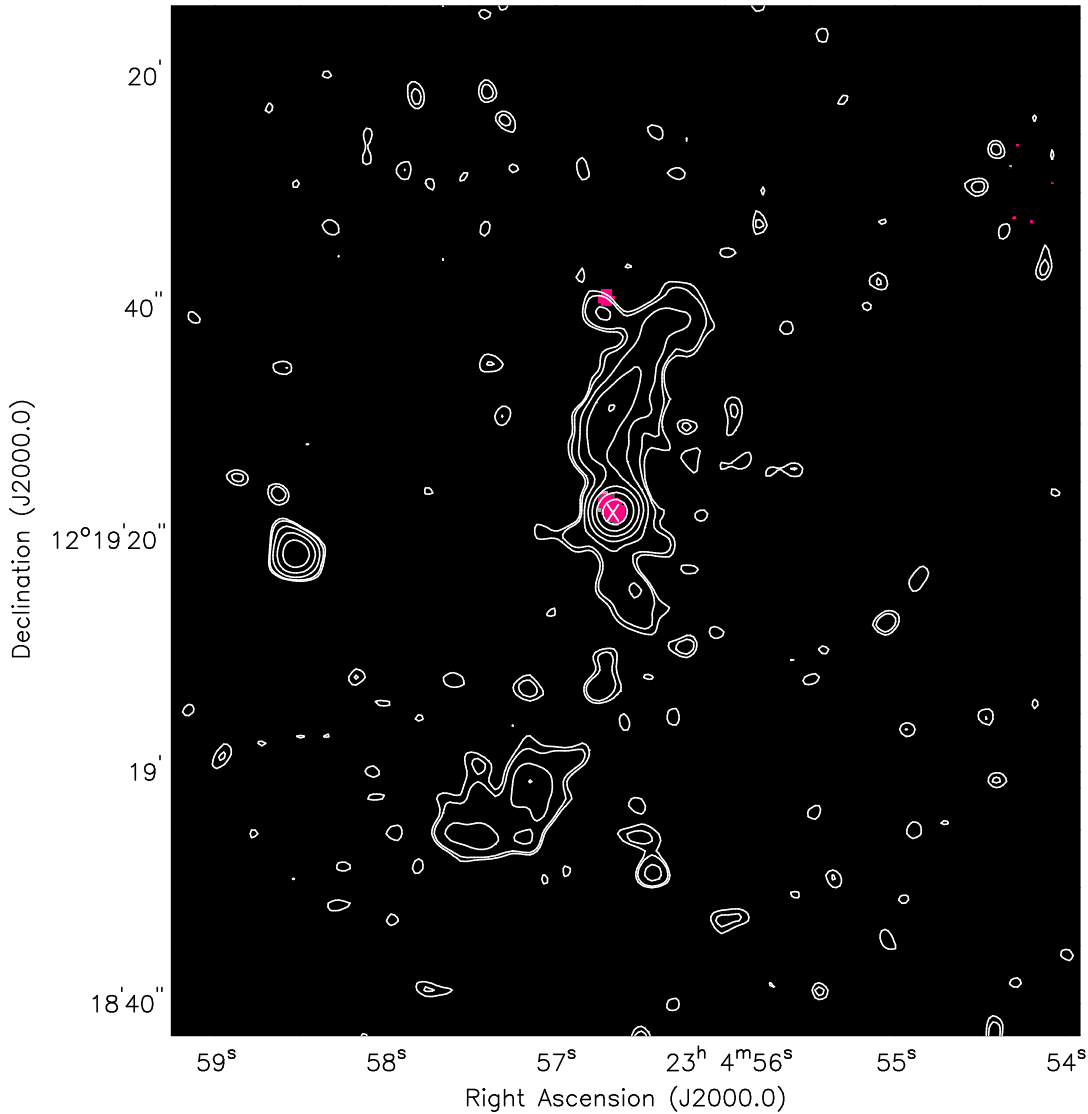}
\caption{Contour (white) image of the 3.5~cm total intensity (Stokes I)
emission from the center of NGC 7479 at 2\farcs2 resolution (C-array),
overlaid on an optical $B$-band image. The nucleus is marked with a white
``X'' and the 22 mag~arcsec$^{-2}$ isophote on the B image is drawn with a
black contour to show the stellar bar. The rms noise is
$13~\mu$Jy~beam$^{-1}$. The contour levels are at $(3, 4, 8, 16, 32,
64, 128)\times11~\mu$Jy~beam$^{-1}$. No polarized emission is detected
at this resolution. The beam is shown at the bottom left
corner.\label{arcsec2fig}}
\end{figure*}

\begin{figure*}[th]
\includegraphics[width=3.5in]{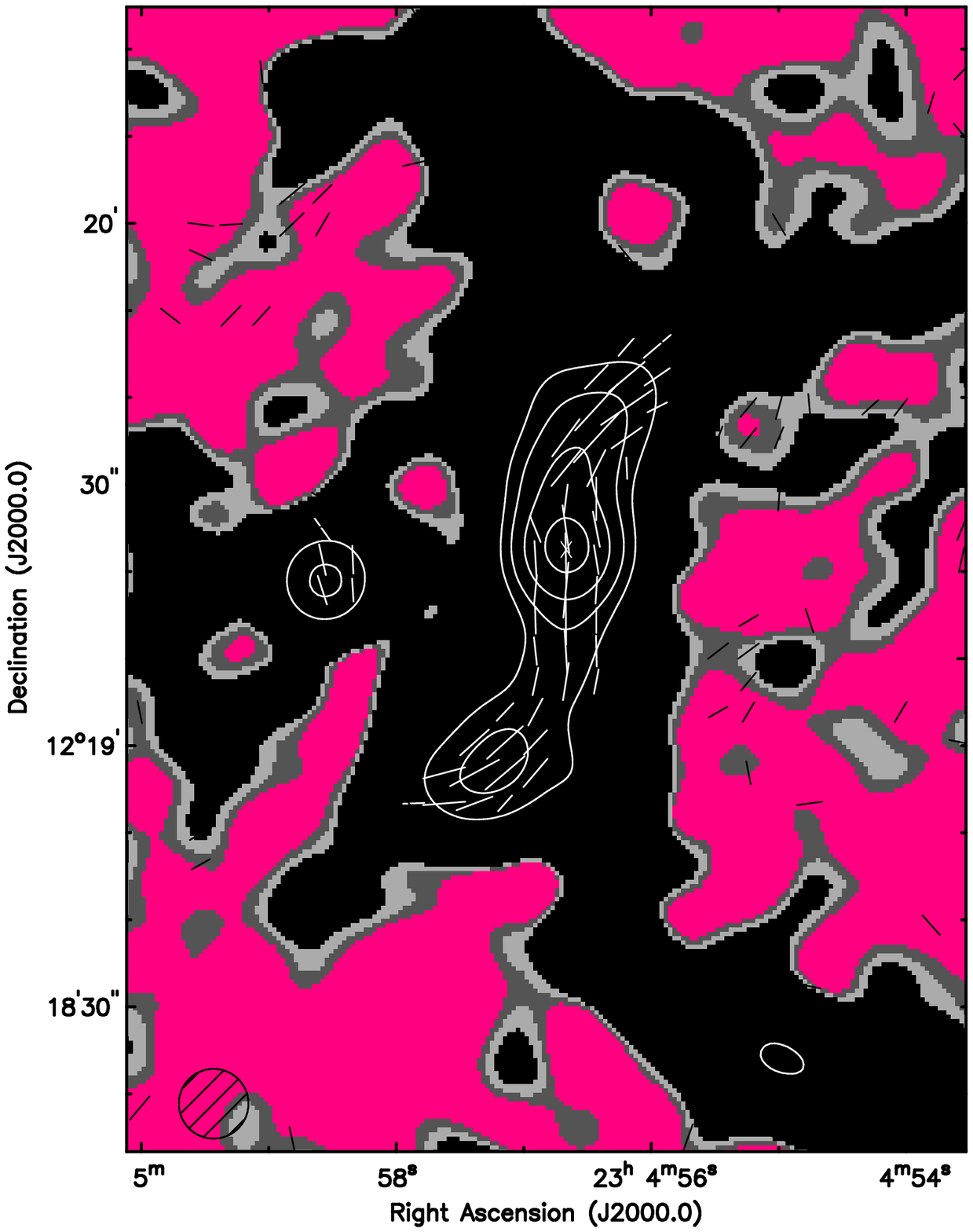}
\hfill
\includegraphics[width=3.5in]{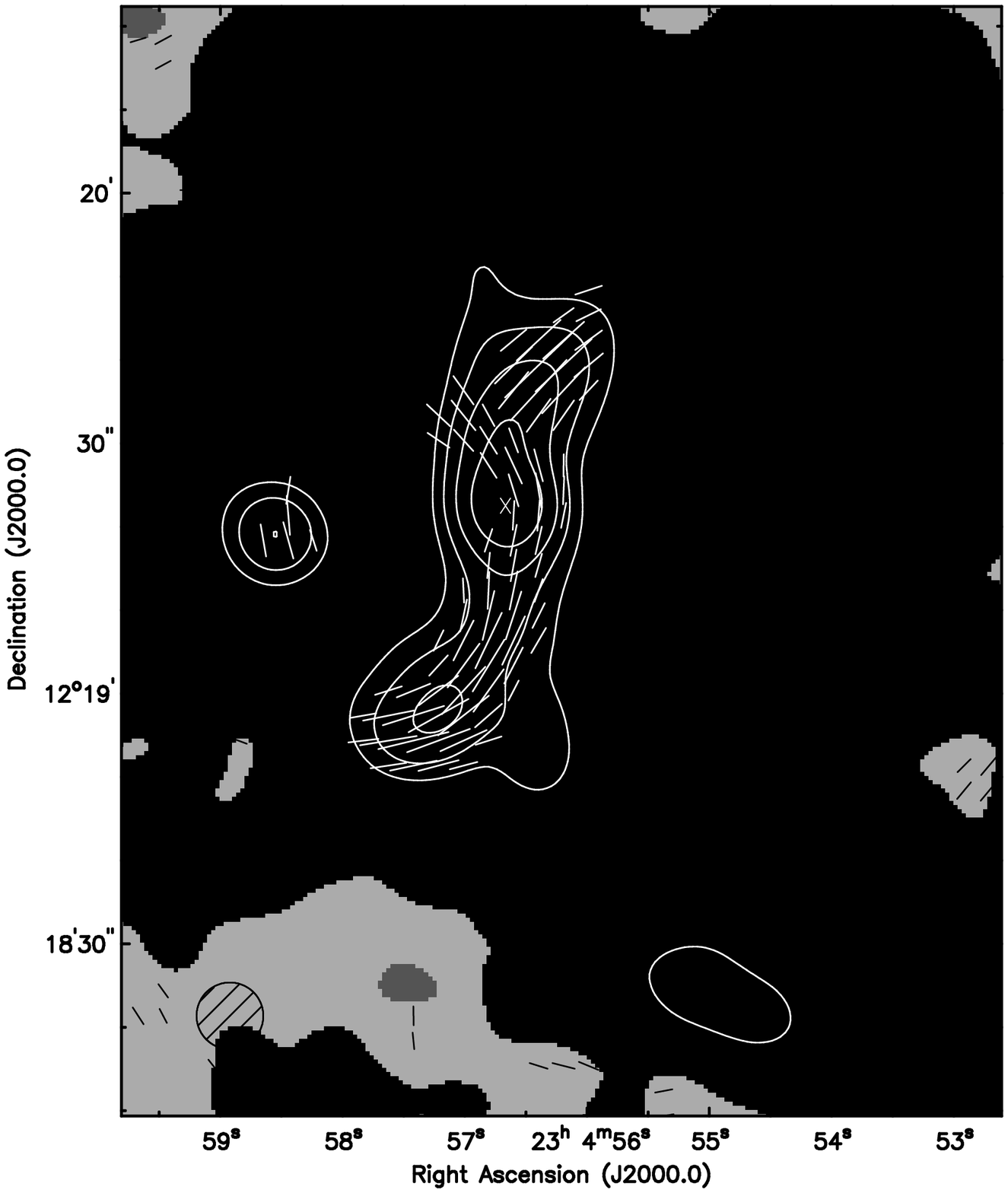}
\caption{a) B-vectors (E-vectors rotated
by 90\degr) overlaid on a grayscale and contour image of the 3.5~cm
total intensity at 8\arcsec\ resolution (D-array). The rms noise is
$11~\mu$Jy~beam$^{-1}$. The contours are at $(2, 4, 8, 16, 32, 64,
128, 256)\times12~\mu$Jy~beam$^{-1}$. The telescope beam is shown at
the bottom left corner. The nucleus is shown with a white cross.
b) B-vectors (E-vectors rotated by 90\degr, not corrected for Faraday
rotation) overlaid on a grayscale and contour image of the 6~cm
total intensity at 8\arcsec\ resolution (C-array). The rms noise is
$9~\mu$Jy~beam$^{-1}$. The contours are at $(2, 4, 8, 16, 32, 64,
128, 256)\times10~\mu$Jy~beam$^{-1}$. The telescope beam is shown
close to the bottom left corner. The nucleus is shown with a white
cross.\label{arcsec8fig}}
\end{figure*}

For maximum sensitivity we used two intermediate frequencies (IFs),
with bandwidths of 50~MHz, and separated by 50~MHz, except in the
case of 18 and 22 cm observations when we used only one IF at each
wavelength. After minimal editing of the visibilities and
calibration, we used the AIPS task {\sc imagr} to Fourier transform
the observed visibilities into brightness distribution maps on the
sky. We applied phase-only self-calibration to the 6~cm C-array
data, resulting in a slight decrease in rms noise. Final cleaning
was commonly performed for 10000 iterations, except in the highest
resolution maps in which we cleaned for 50000 iterations. Varying the
value of the {\sc robust} parameter, maps with different resolutions
were made. The best resolution of the C-array data is 2\farcs2
(synthesized half-power beamwidth) at 3.5~cm and about 4\arcsec\ at
6~cm. These maps were subsequently smoothed to circular beams of
6\arcsec, 8\arcsec, and 12\arcsec. The D-array data at 3.5~cm give a
best resolution of 8\arcsec, with an rms noise about half that of
the smoothed map obtained from the C-array data. We did not find
substantial improvement in maps made from combined C and D array
$uv$ data, and chose to use the 3.5~cm images made with data from
separate configurations. No correction for the primary beam
attenuation was made. The D-array data at 6~cm from \citet{beck02},
obtained from 30 minutes of total observing time, are not sensitive
enough for the purposes of this paper. The more recent D-array data
at 6~cm could not be used due to a severe hardware problem during
the observations, so that only the C-array data at 6~cm were used
for this paper.

We made maps in the Stokes parameters I, Q, and U. The largest
structures visible in our observations are about 3\arcmin\ at 3.5~cm
and 5\arcmin\ at 6~cm. Hence our maps can be used reliably to
study the spectral index only in the central bright parts, including
the jet-like structure. It is unlikely that any large-scale
structure is missing in the Stokes parameters Q and U, since the
polarization angle changes throughout the galaxy. We made maps of
the linearly polarized intensity (PI) and the polarization angle
(PA) from the Q and U maps, adding 90\degr\ to obtain B-vectors. The
positive bias in PI was corrected with the {\sc polco} routine in
AIPS.

To map the largest structures, 3.6~cm observations of NGC~7479 were
performed in February 2003 under excellent weather conditions with
the Effelsberg single-dish telescope using the 8.35~GHz receiver
with 1.1~GHz bandwidth and a beamsize of 84\arcsec. A field of
$10\arcmin \times 10\arcmin$ was scanned alternately in Right
Ascension and in Declination. 11 coverages were combined using the
standard NOD2 software package. The final maps in Stokes I, Q, and U
were smoothed to 90\arcsec\ resolution. The rms noise is
$\simeq400~\mu$Jy~beam$^{-1}$ in total intensity (limited by
scanning effects) and $70~\mu$Jy~beam$^{-1}$ in Q and U. A
combination of VLA and Effelsberg 3~cm data was not attempted as we
concentrate on the jet properties in this paper.

\section{RESULTS}

\subsection{Total Intensity and Spectral Index}
\label{s:maps}

\begin{figure}[th]
\centering
\includegraphics[width=3.55in]{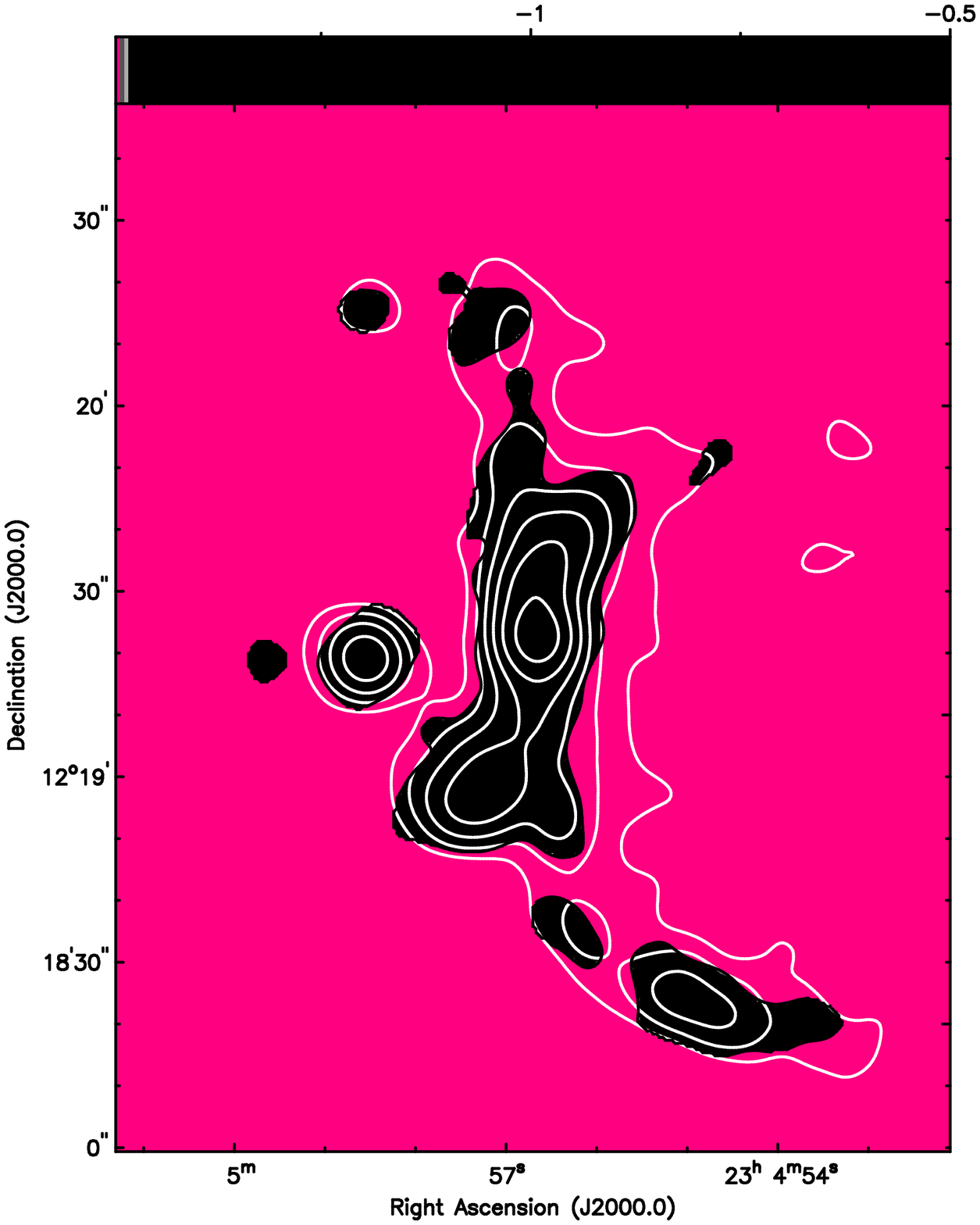}
\caption{Grayscale image of the spectral index between 3.5 and
6~cm at 8\arcsec\ resolution, overlaid by contours of 6~cm total
intensity. The contours are at $(2, 4, 8, 16,
32)\times50~\mu$Jy~beam$^{-1}$.\label{spindexat8}}
\end{figure}

\begin{figure*}[th]
\centering
\includegraphics[width=6.5in]{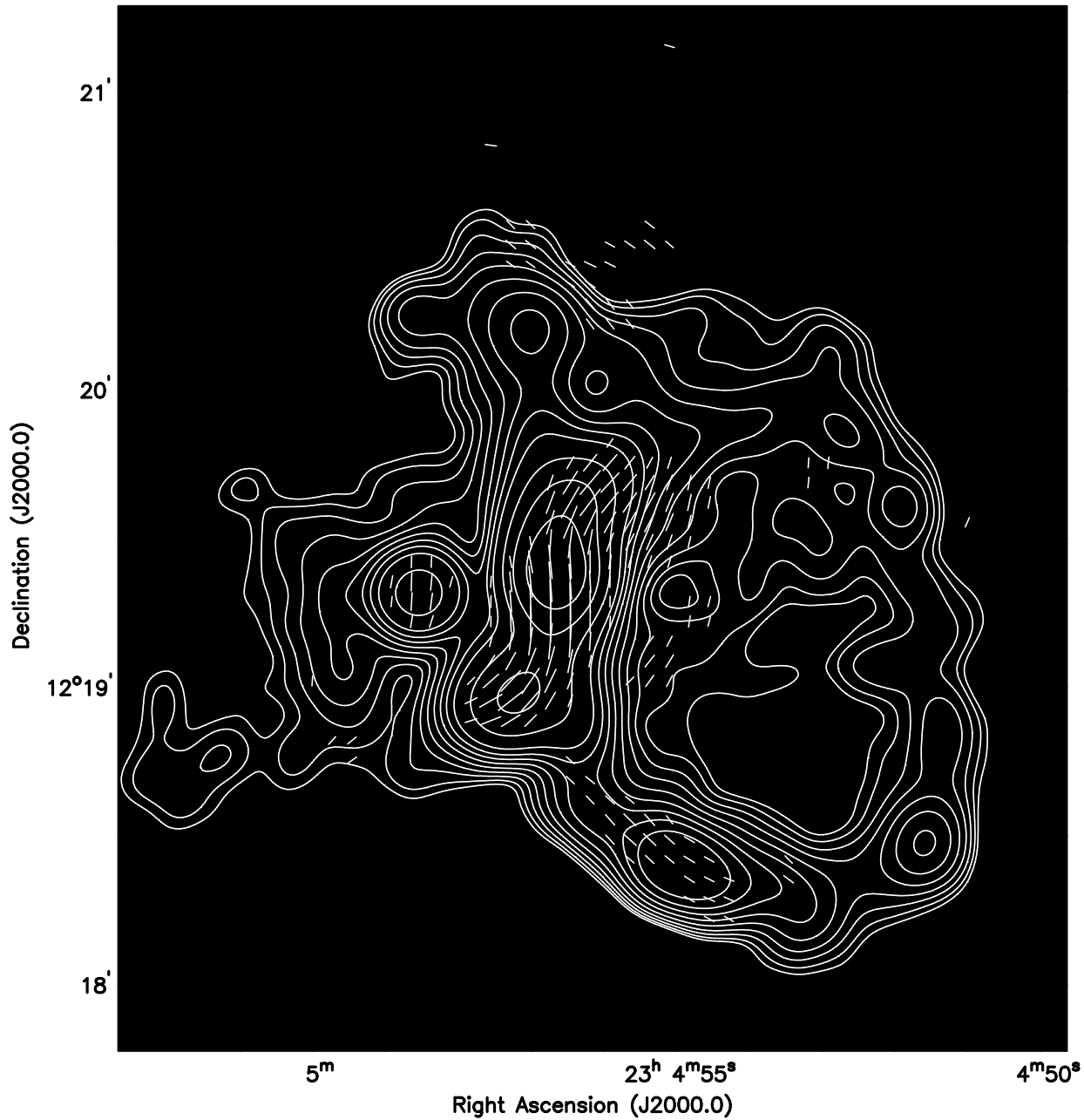}
\caption{B-vectors (E-vectors rotated by
90\degr) and contour image of the 3.5~cm total intensity at
12\arcsec\ resolution (D-array), overlaid on an unpublished {\em
Spitzer\/} grayscale image of 3.6~$\mu$m infrared emission at
2\arcsec\ resolution. The rms noise is $10~\mu$Jy~beam$^{-1}$. The
contours are at $(1, 1.4, 2, 2.8, 4, 5.7, 8, 11, 16, 32, 64, 128,
256)\times20~\mu$Jy~beam$^{-1}$.\label{arcsec12fig}}
\end{figure*}

\begin{figure*}[th]
\centering
\includegraphics[width=6.5in]{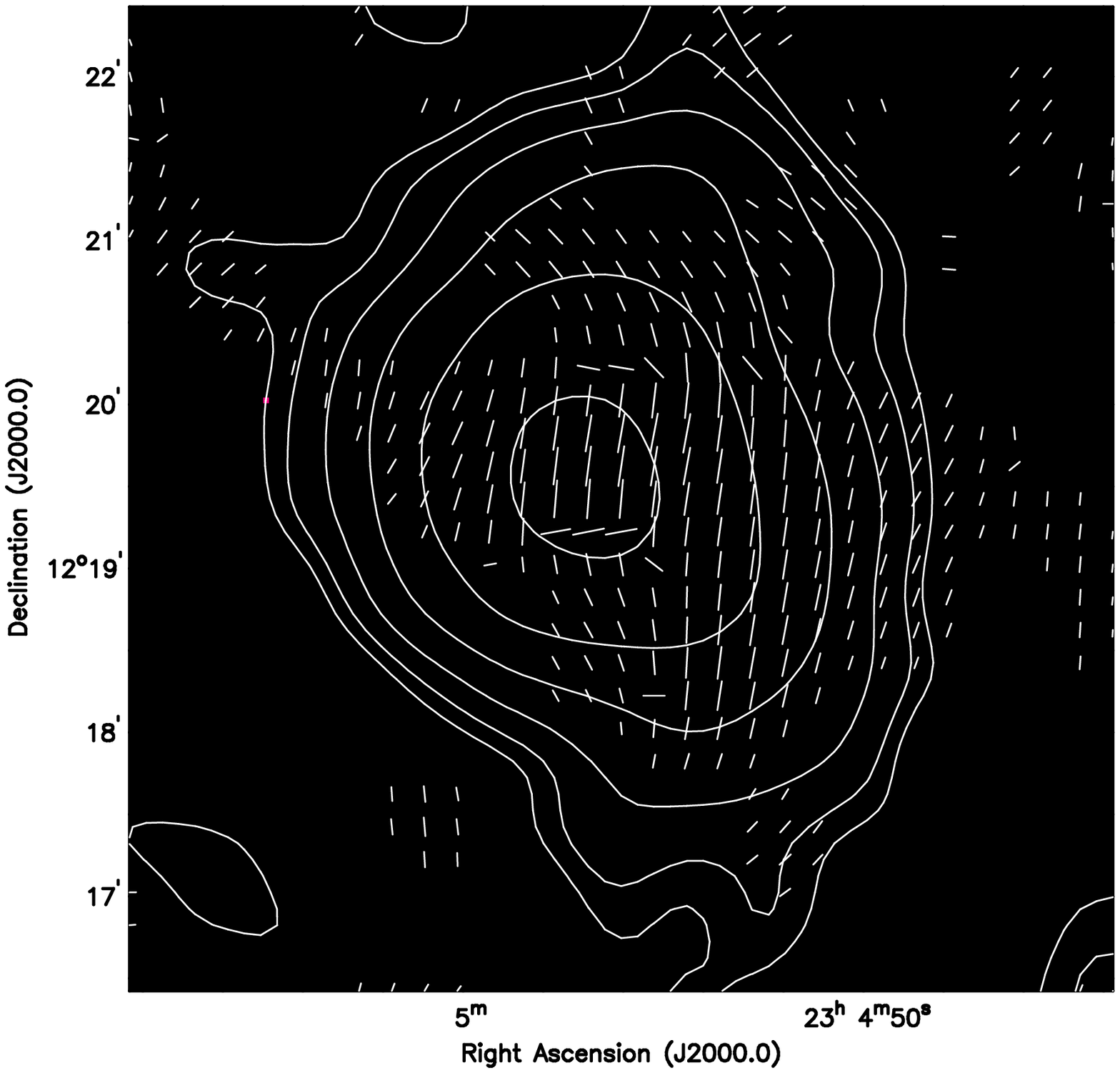}
\caption{Effelsberg B-vectors (E-vectors
rotated by 90\degr) and contour image of the 3.6~cm total intensity
at 90\arcsec\ resolution, overlaid on a grayscale Digital Sky Survey
optical image. The rms noise is $\simeq 400~\mu$Jy~beam$^{-1}$. The
contours are at $(1, 2, 4, 8, 16, 32)\times400~\mu$Jy~beam$^{-1}$.\label{36cmarcsec90fig}}
\end{figure*}

A VLA contour map of the total radio intensity at 2\farcs 2 resolution
at 3.5~cm is shown in Figure~\ref{arcsec2fig}, and maps of the total
radio intensity at 8\arcsec\ resolution at 3.5 and 6~cm are shown in
Figure~\ref{arcsec8fig}. The ``S''-shaped, double-jet feature is the
dominant structure. There is little indication (lack of detailed
spatial correlation with bar structure in high resolution images) that
a considerable fraction of the radio continuum emission emerges from
the bar region (at $+6\degr$ position angle), although we cannot
exclude this possibility. The emission along the jet is asymmetric,
with the northern side being stronger in total intensity by about 30\%
in the 8\arcsec\ maps at 3.5 and 6~cm. At the highest resolution
(Figure~\ref{arcsec2fig}) the shape of the northern jet is also more
regular and narrower than the southern jet which breaks into diffuse
clouds separated by apparent gaps due to the low signal-to-noise at
this resolution. The full width at half power of the northern jet is
2\farcs7. Its physical width, corrected for beam smearing, is about
250~pc. At 8\arcsec\ resolution (Figure~\ref{arcsec8fig}) the southern
jet is also continuous, but with an intensity minimum at about
17\arcsec\ from the center. The projected jet lengths in
Fig.~\ref{arcsec8fig} are 5.5~kpc and 6.5~kpc in the north and south,
respectively.

Radio emission from the bar and the adjacent spiral arms is also
detected (Fig.~\ref{arcsec8fig}), but it is
much weaker than the emission from the jet. Possibly a fraction of
the radio continuum emission along the inner jet comes from the bar
component, related to amplified magnetic fields in the shocks of the
bar \citep{rbeck05}. We tried to subtract this component using
Spitzer observations and a known far-IR--radio continuum correlation,
but unfortunately the resolution of the far-IR Spitzer observations
is too low (close to 20\arcsec\ at 70~$\mu$m) and there is a strong
nuclear point source with extended diffraction structure which is
difficult to remove and which makes the estimation of a bar
component even harder.

The spectral index map at 8\arcsec\ resolution in
Figure~\ref{spindexat8} shows that the spectral index between 3.5
and 6~cm is close to $-0.5 \pm 0.03$ near the nucleus and $-0.8 \pm
0.04$ along the jet. (Although the cited uncertainties are valid for
the absolute values of the spectral indices, the uncertainty in
relative values between different locations in the image is smaller,
since all the pixels in a given image are subject to common
calibration uncertainties. The uncertainty in the relative spectral
index between adjacent regions in the image is determined by the rms
noise values in the individual images, and is $\approx$~$\pm 0.01$
near the nucleus and $\approx$~$\pm 0.03$ in the jet.) As thermal
emission from the nucleus and the jet is probably small, the above
values can be taken as the synchrotron spectral index. No
significant steepening is observed along the jet, which indicates
fast propagation of cosmic rays along the jet. The mean value of the
spectral index between 3.5 and 6~cm in the southern spiral arm is
around $-0.75$. Assuming a typical thermal fraction of 25\% at 6~cm,
the synchrotron spectral index is $\simeq -1.0$.

The image at 12\arcsec\ resolution (Figure~\ref{arcsec12fig}) has
high sensitivity to diffuse extended emission. The two
optical/near-infrared spiral arms emerging from the ends of the bar,
specifically the southwestern arm, are also seen in the total radio
intensity map. Clouds of diffuse radio emission west of the bar
(between the bar and the outer spiral arm) and also east of the bar
indicate a reservoir of relativistic electrons and magnetic fields.
In the Effelsberg 3.6~cm map (Figure~\ref{36cmarcsec90fig}) and in
the VLA maps at 18~cm and 20~cm wavelengths \citep{beck02} diffuse
emission in total intensity maps can be followed out to about 23~kpc
projected radius.

\subsection{Polarized Intensity}
\label{s:pol}

\begin{figure*}[th]
\includegraphics[width=3.5in]{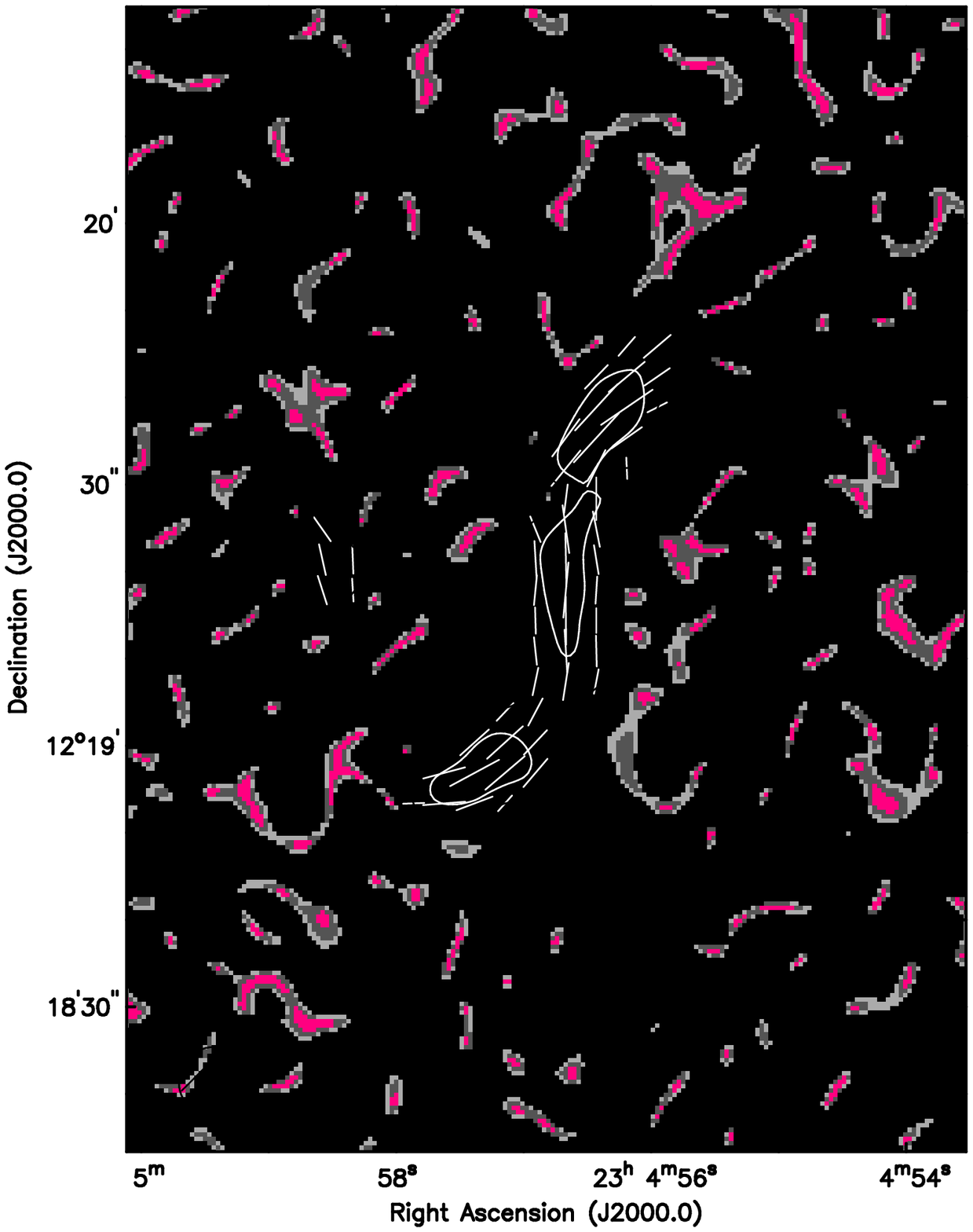}
\hfill
\includegraphics[width=3.5in]{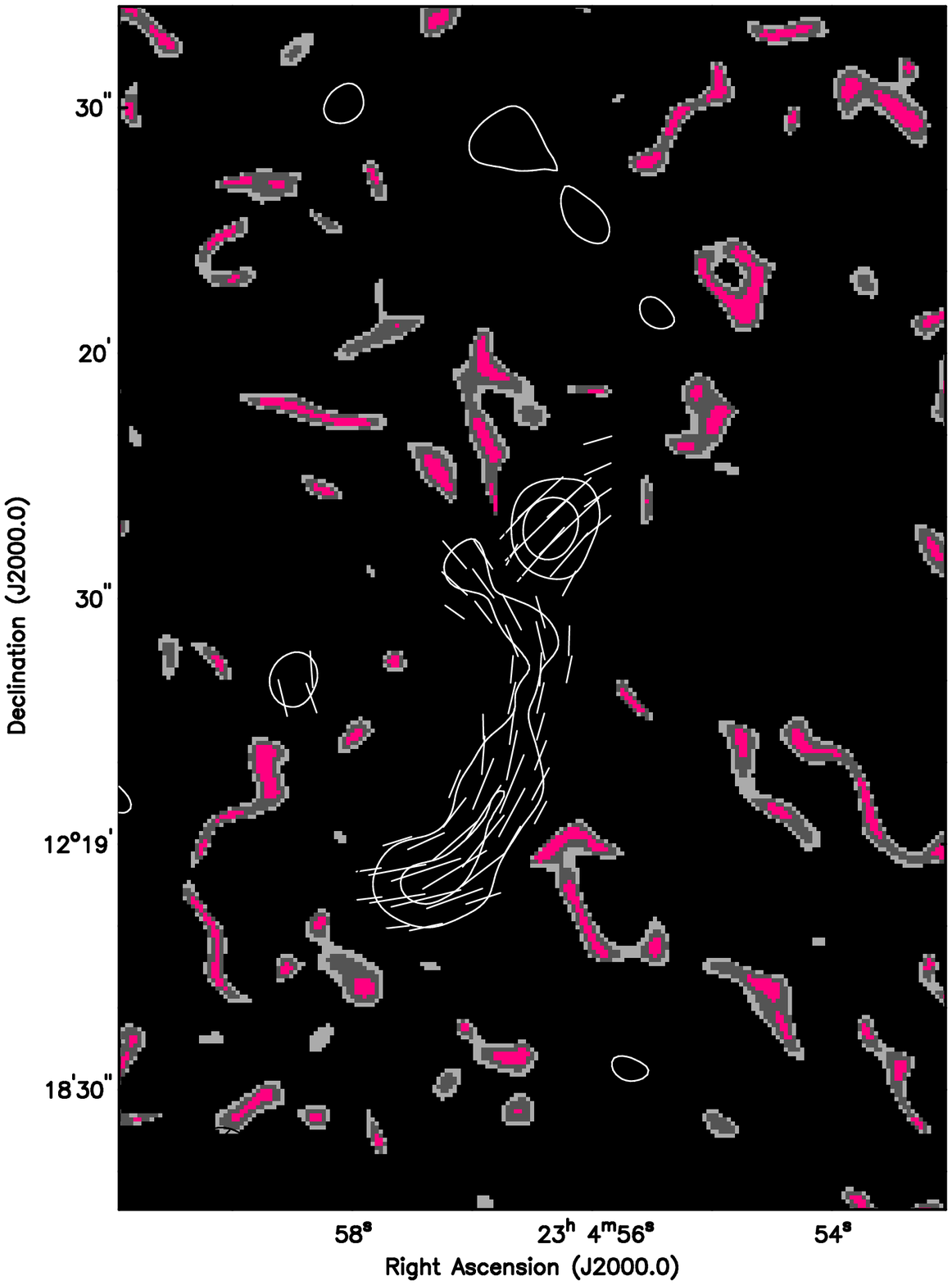}
\caption{a) B-vectors (E-vectors rotated
by 90\degr) overlaid on a contour and grayscale image of polarized
intensity of the 3.5~cm radio continuum emission at 8\arcsec\
resolution (D-array). The contours are at $(2,
4)\times12~\mu$Jy~beam$^{-1}$. The rms noise is
$11~\mu$Jy~beam$^{-1}$. b) B-vectors (E-vectors rotated by 90\degr,
not corrected for Faraday rotation) overlaid on a contour and
grayscale image of polarized intensity of the 6~cm radio continuum
emission at 8\arcsec\ resolution (C-array). The rms noise is
$9~\mu$Jy~beam$^{-1}$. The contours are at $(2, 4,
8)\times10~\mu$Jy~beam$^{-1}$. Beams are shown by the hatched circle
near the bottom left corners.\label{arcsec8polfig}}
\end{figure*}

We present 8\arcsec\ resolution maps of the polarized emission at
3.5 and 6~cm in Figure~\ref{arcsec8polfig}. The strongest
polarized emission emerges from blobs at the jet ends. Polarized
intensity appears to be somewhat higher at 6~cm than at 3.5~cm.
However, as the signal-to-noise ratio at 3.5~cm is worse than at
6~cm, the intensity differences are hardly significant (while the
angle difference is significant, see below).

The position angles of the B-vectors at 3.5~cm remain almost
constant (140\degr\ -- 180\degr) along the jet, but decrease to
$\approx$~90\degr\ towards the southern end of the jet, and to
$\approx$~45\degr\ towards the northern end, giving an indication of
bending of the jet at its ends.

The polarized emission at 6~cm in the outer northern jet emerges
mostly from a source near the end of the jet at RA$_{2000}$=
23h04m56\fs3, DEC$_{2000}=12\degr 19\arcmin 39\arcsec$ which is
barely visible in the maps of total emission. No source is seen at
this location in the optical and infrared maps, only a dust lane
extending towards south-southwest. The polarized emission at 3.5~cm
is elongated along the outer northern jet, with only a small gap
towards the inner jet. The B-vectors at 6~cm
(Fig.~\ref{arcsec8polfig}b) in the northern jet, south of the
polarized source, turn towards the northeast due to strong Faraday
rotation between 3.5~cm and 6~cm, while Faraday rotation in the
polarized source is smaller (see Section~\ref{s:rm}). Hence the gap
at 6~cm between the polarized source and the inner jet is due to
Faraday depolarization.

The degree of polarization is remarkably constant along the jet,
with values between 6\% and 8\% at 3.5~cm. Lower values near the
nucleus are caused by the strong unpolarized emission from the
nucleus. The degree of polarization at 6~cm is similar to that at
3.5~cm in the outer jet, but lower in the inner jet due to Faraday
depolarization.

The B-vectors in Fig.~\ref{arcsec8polfig} show an ordered\footnote{Note
that magnetic fields observed in linearly polarized
emission can be either {\em coherent} (i.e., preserving their
direction within the telescope beam) or {\em incoherent} (i.e., with
multiple field reversals within the beam). To distinguish between
these two components, additional Faraday rotation data are needed.
For the sake of simplicity, the fields observed in polarization are
called ``ordered'' throughout this paper if Faraday rotation is not
measured. If Faraday rotation is measured and it is significant,
``regular field'' is used.} pattern and follow the jet axis. The
B-vectors at 3.5~cm (Fig.~\ref{arcsec8polfig}a) are
Faraday-rotated by less than 10\degr\ (see Sect.~\ref{s:rm}) and
hence delineate the orientation of the ordered magnetic field. We
conclude that the field precisely follows the S-shape of the jet
which is important for understanding its origin
(Section~\ref{s:field}).

\begin{figure*}[th]
\centering
\includegraphics[width=6.5in]{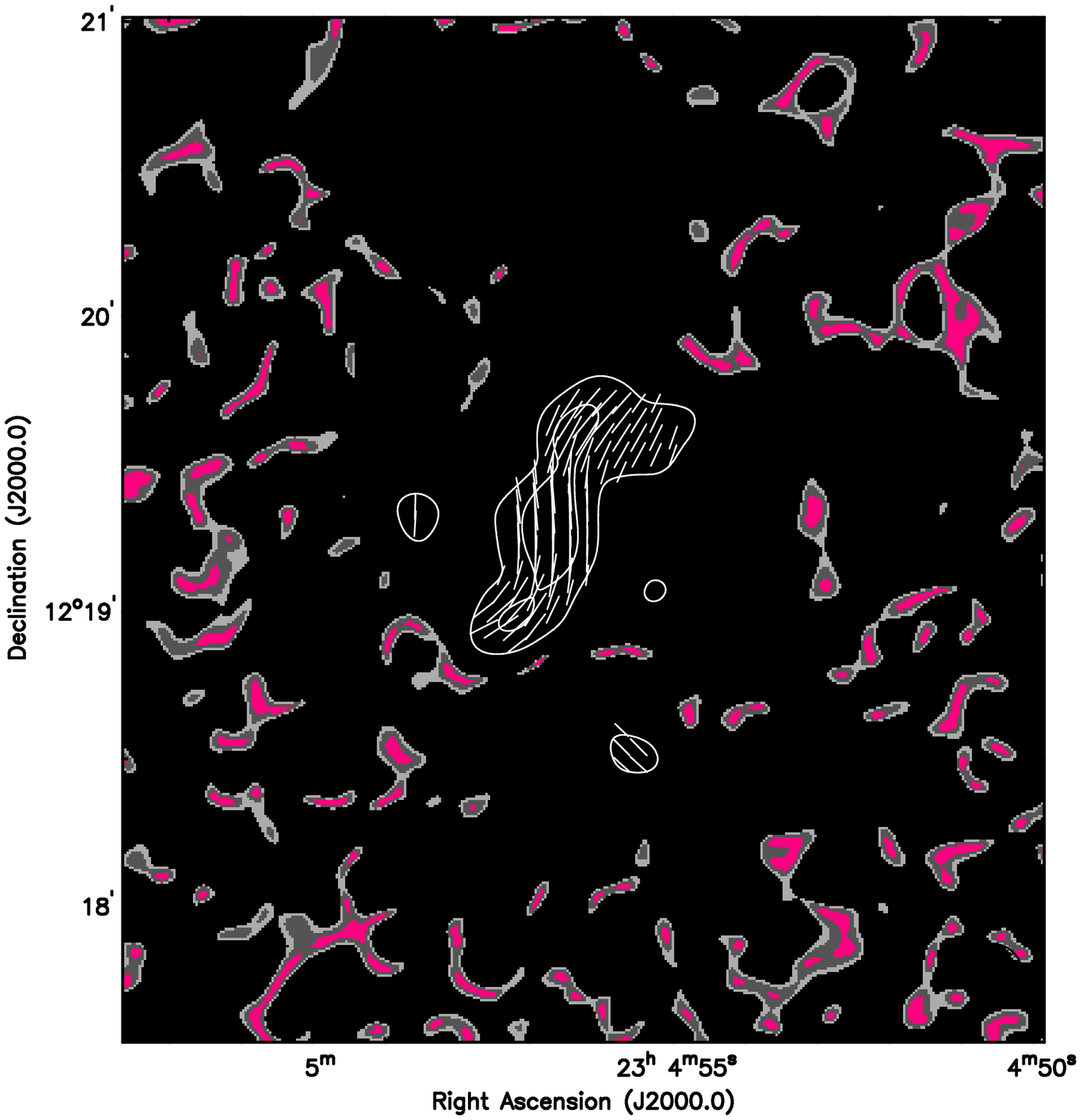}
\caption{B-vectors (E-vectors rotated by
90\degr) and contour image of the 3.5~cm polarized intensity at
12\arcsec\ resolution (D-array). The rms noise is
$9~\mu$Jy~beam$^{-1}$. The contours are at $(2, 4,
8)\times10~\mu$Jy~beam$^{-1}$.\label{arcsec12polfig}}
\end{figure*}

Polarized emission is observed also outside the jet. The elongated
feature at the top of Fig.~\ref{arcsec8polfig}b is located in the
interarm region between the northern spiral arms and the northern
part of the wrapping western arm, and is highly polarized ($\approx
50\%$). Interarm polarization is typical for spiral galaxies
\citep{beck05}. Figure~\ref{arcsec12polfig} shows that the interarm
polarization extends to the edge of the optically visible disk of
the galaxy. In contrast, the southern spiral arm shows polarization
{\em on\/} the arm with well-aligned B-vectors and lower degrees of
polarization of 10\%--20\% (Fig.~\ref{arcsec12polfig}). This could
be due to compression as the spiral arm is sweeping the ISM of the
galaxy, as suggested by the merger models of \citet{laine99}. The
western continuation of the arm and the bar itself do not show any
significant polarized emission.

The diffuse clouds of total emission west and east of the bar
(Fig.~\ref{arcsec12fig}) are polarized at 3.5~cm with a well-ordered
magnetic field oriented parallel to the jet axis
(Fig.~\ref{arcsec12polfig}). These features are also seen in the
6~cm polarization map at 30\arcsec\ resolution \citep{beck02}.
Even with the low angular resolution of our Effelsberg map
(Fig.~\ref{36cmarcsec90fig}), in which depolarization within the beam
is strong, these regions are still weakly polarized, with similar
field orientations as in Fig.~\ref{arcsec12polfig}.

\begin{figure*}[th]
\includegraphics[width=3.5in]{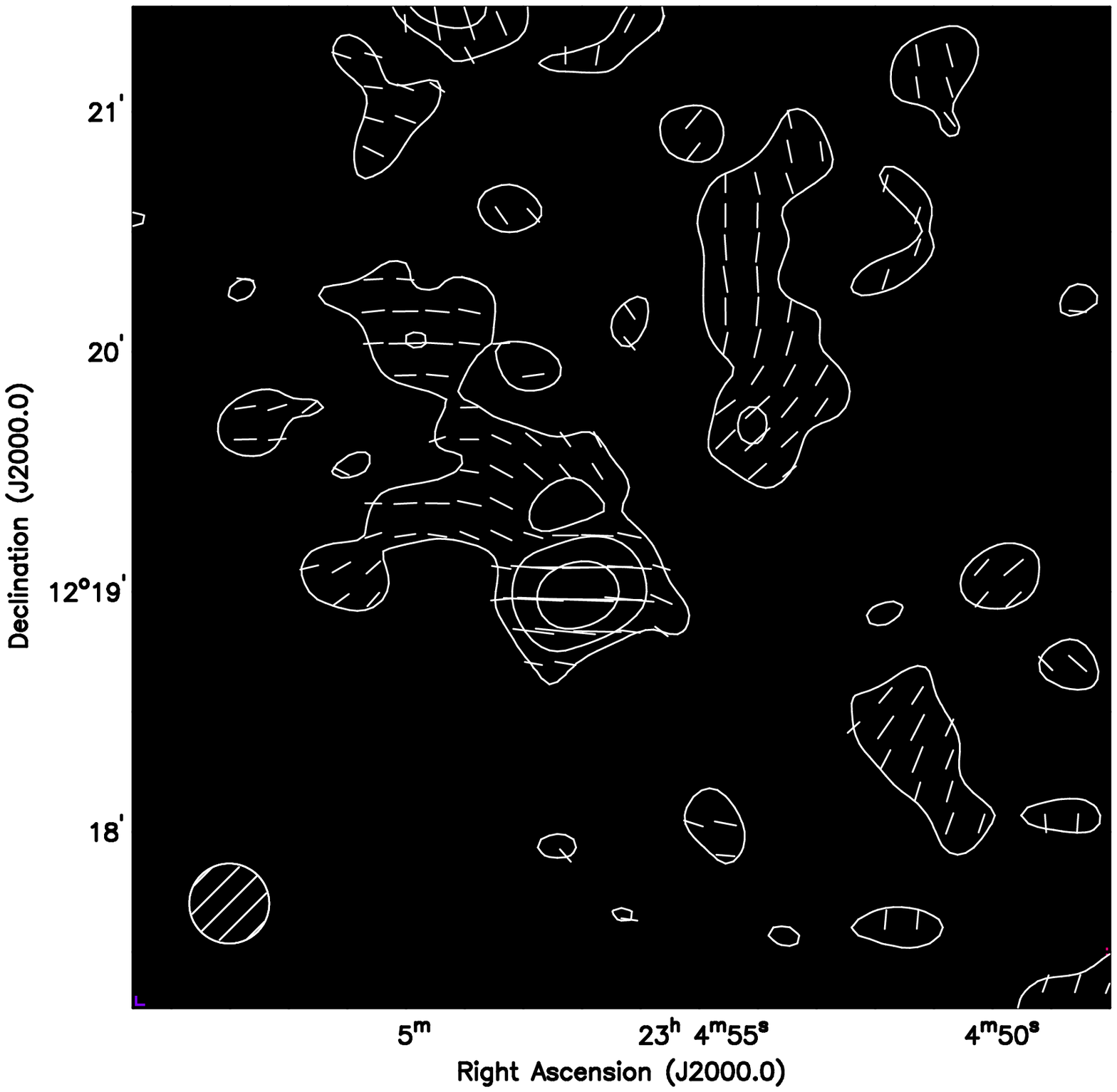}
\hfill
\includegraphics[width=3.5in]{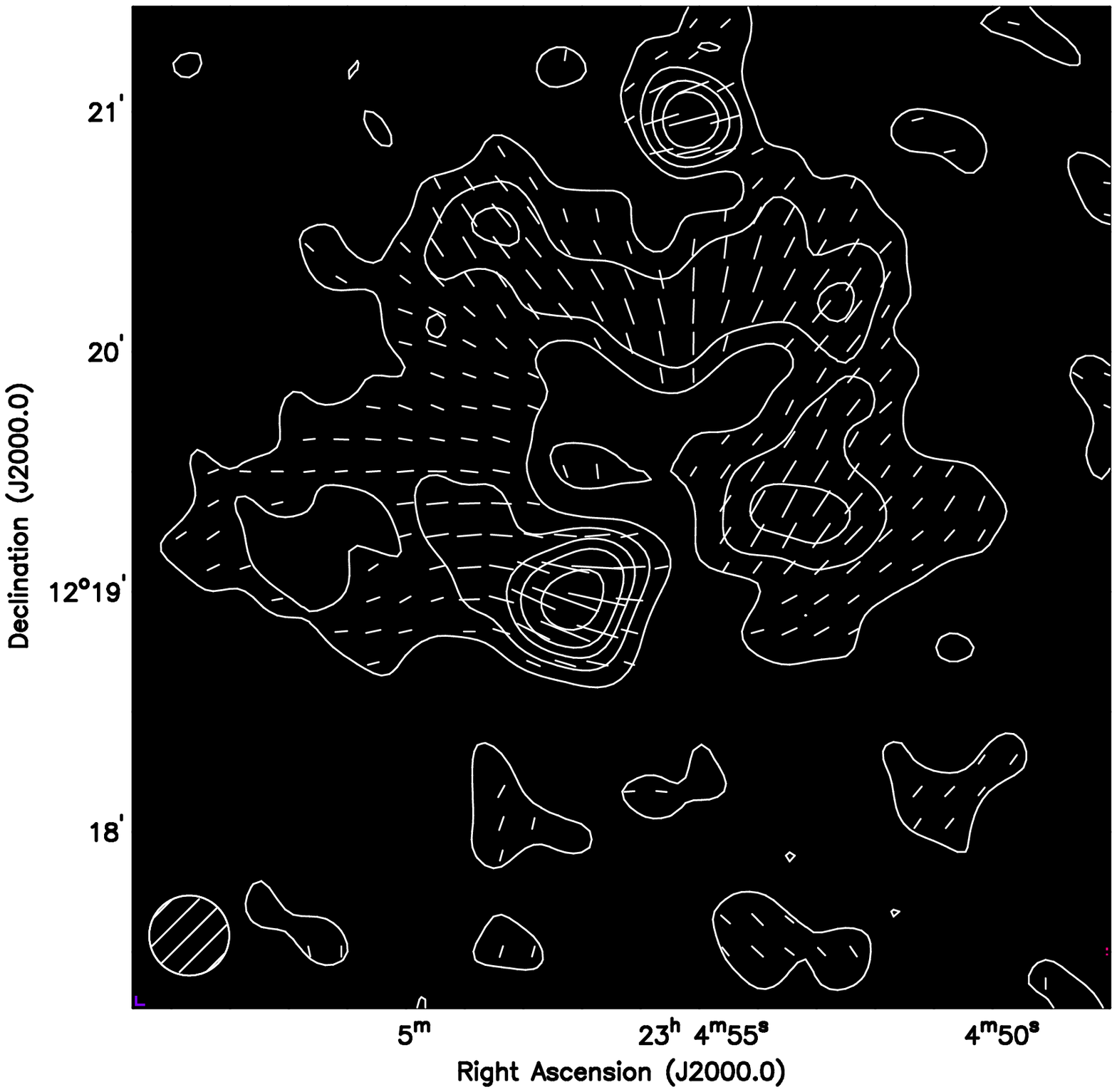}
\caption{a) B-vectors (E-vectors rotated
by 90\degr, not corrected for Faraday rotation) and a contour image
of polarized intensity of the radio continuum emission at 18~cm  and
b) at 22~cm overlaid on a grayscale Digital Sky Survey optical
image, both at 20\arcsec\ resolution. The rms noises are
$40~\mu$Jy~beam$^{-1}$ and $20~\mu$Jy~beam$^{-1}$, respectively. The
contours are at $(2, 4, 8)\times40~\mu$Jy~beam$^{-1}$ in a) and $(1,
2, 3, 4, 6)\times45~\mu$Jy~beam$^{-1}$ in b).\label{arcsec20polfig}}
\end{figure*}

At 18~cm and 22~cm (Figure~\ref{arcsec20polfig}) polarized emission
is detected from the southern jet only (with $p\simeq5\%$
fractional polarization), while the northern jet is completely
depolarized. Such asymmetric depolarization is well known in radio
galaxies as the {\em Laing-Garrington effect\/} \citep{garr88} and
suggests that the northern jet is located behind the galaxy disk
which depolarizes the emission, while the southern jet is in front
of the disk (Section~\ref{s:dp}).

The diffuse clouds east and west of the bar are polarized at 22~cm
by $p\simeq10\%$. They are hardly detected at 18~cm because
the signal-to-noise ratio is too low. Field strengths and ionized
gas density are low in the outer disk, so that Faraday
depolarization is smaller than in the inner parts. The overall
distribution of diffuse polarized intensity at 22~cm is asymmetric
along the major axis of the inclined disk (position angle of
22\degr). This phenomenon of asymmetric Faraday depolarization
around 20~cm wavelength is observed in many spiral galaxies,
e.g., in NGC~6946 \citep*{beck07}.

\subsection{Magnetic Field Strengths in the Disk and the Jet}
\label{s:strength}

The strengths of the total magnetic field $B_{\rm t}$ and its
ordered component in the sky plane can be derived from the total
synchrotron intensity and its degree of linear polarization at
8\arcsec\ resolution (Sect.~\ref{s:pol}), assuming equipartition
between the energy densities of the magnetic field and the total
cosmic rays, a value for the ratio $R$ between the number densities
of cosmic-ray protons and electrons, the pathlength $L$ through the
synchrotron-emitting medium, and the synchrotron spectral index
$\alpha_n$ \citep{bk05}. The equipartition strengths of the ordered
field, either regular or incoherent, are always lower limits due to
the limited resolution (beam depolarization) and Faraday
depolarization effects.

The total field strength $B_{\rm t}$ in the southern spiral arm
(assuming $R=100$, $L=1$~kpc for the full disk thickness and
$\alpha_{\rm n}=-1.0$, see Fig.~\ref{spindexat8}) is
$\simeq18~\mu$G, and that of the ordered field in the sky plane (not
corrected for inclination) is $\simeq7~\mu$G. The strength of the
random field B$_{\rm r}$ is $\simeq17~\mu$G. Outside the bar and
spiral arms the average total field strength is $\simeq11~\mu$G and
that of the ordered field in the sky plane is $\simeq8~\mu$G. The
strength of the random field B$_{\rm r}$ is $\simeq7~\mu$G. Such
values are typical for regions between spiral arms  in galaxies
\citep{beck05}.

The total equipartition field strength in the jet, assuming
$R_1=100$ (a relativistic proton-electron plasma), $L=250$~pc
(Sect.~\ref{s:maps}), and $\alpha_{\rm n}=-0.8$
(Fig.~\ref{spindexat8}), varies from $B_{\rm t}\simeq35~\mu$G in the
inner jet to $\simeq40~\mu$G in the outer jet, while the strength
of the component of the ordered field in the sky plane is
$\simeq8~\mu$G and almost constant. In a $40~\mu$G magnetic field,
the synchrotron lifetime of cosmic-ray electrons emitting at 6~cm is
about $10^6$~years. If the jet is composed of a relativistic
positron-electron plasma ($R_2=0$), all equipartition field
strengths are smaller by a factor of $(R_1+1)^{1/(3-\alpha_{\rm n})}
\simeq 3.4$ \citep{bk05}.

These values are based on our radio data at 8\arcsec\ (1.2~kpc)
resolution, while the total intensity map at higher resolution
(Figure~\ref{arcsec2fig}) reveals a width of the jet of $\simeq
250$~pc, so that the internal field structure of the jet is not
resolved and the polarized intensities suffer from depolarization
within the telescope beam. Hence, the equipartition value for the
ordered field is a lower limit.

\subsection{Faraday Rotation: Regular Magnetic Fields}
\label{s:rm}

\begin{figure}[th]
\centering
\includegraphics[width=3.5in]{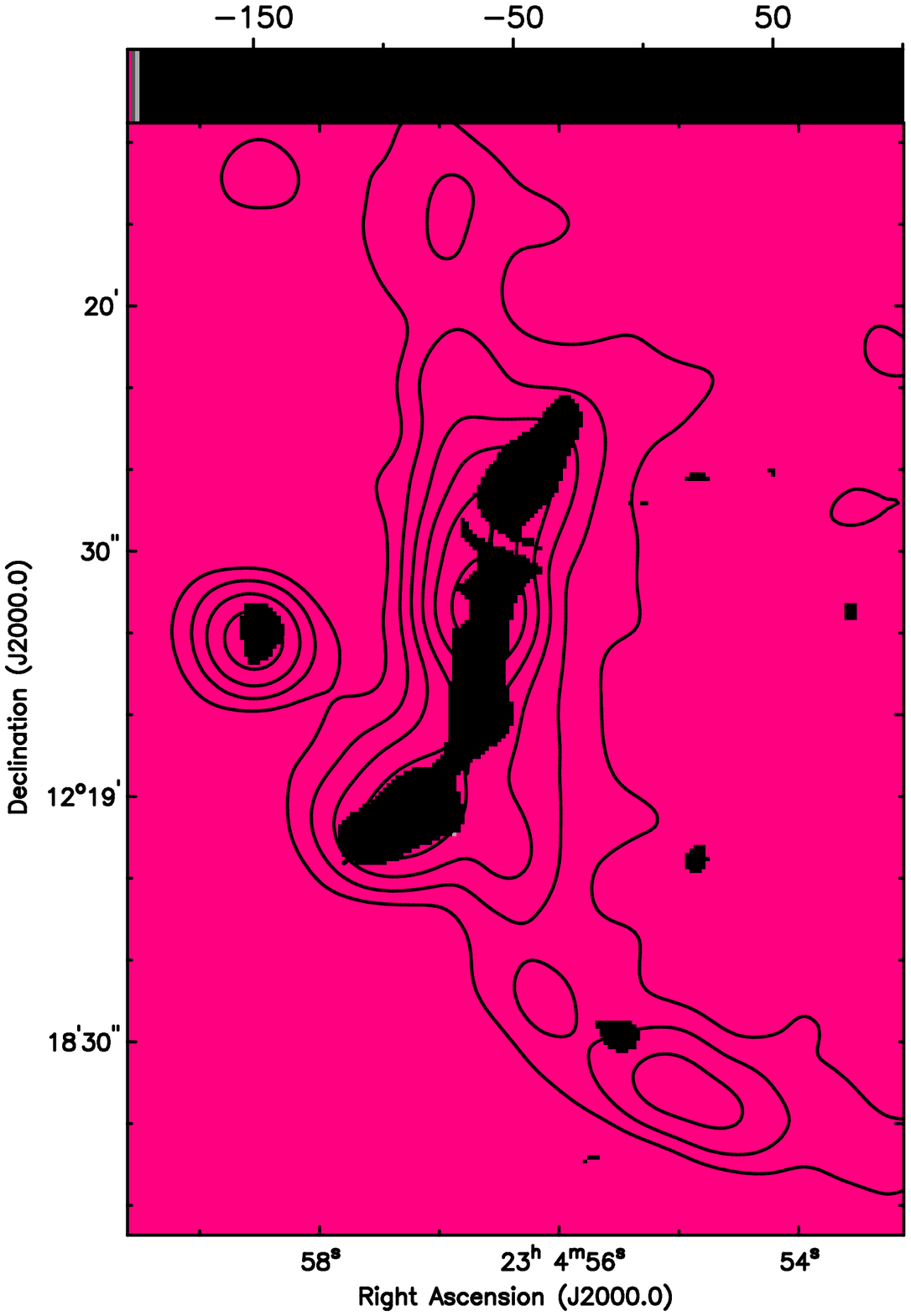}
\caption{Rotation measure (RM) map between 3.5 and 6~cm of
NGC~7479 at 8\arcsec\ resolution in grayscale, overlaid by contours
of total 6~cm intensity. The contour levels are at $(2, 4, 8, 16,
32, 64)\times50~\mu$Jy~beam$^{-1}$. The location of the slices taken
along the inner jet and the outer southern jet
(Figure~\ref{rmslice}) are indicated by thick black lines.}
\label{rm3.5cm6}
\end{figure}

\begin{figure*}[th]
\includegraphics[width=0.35\textwidth,angle=270]{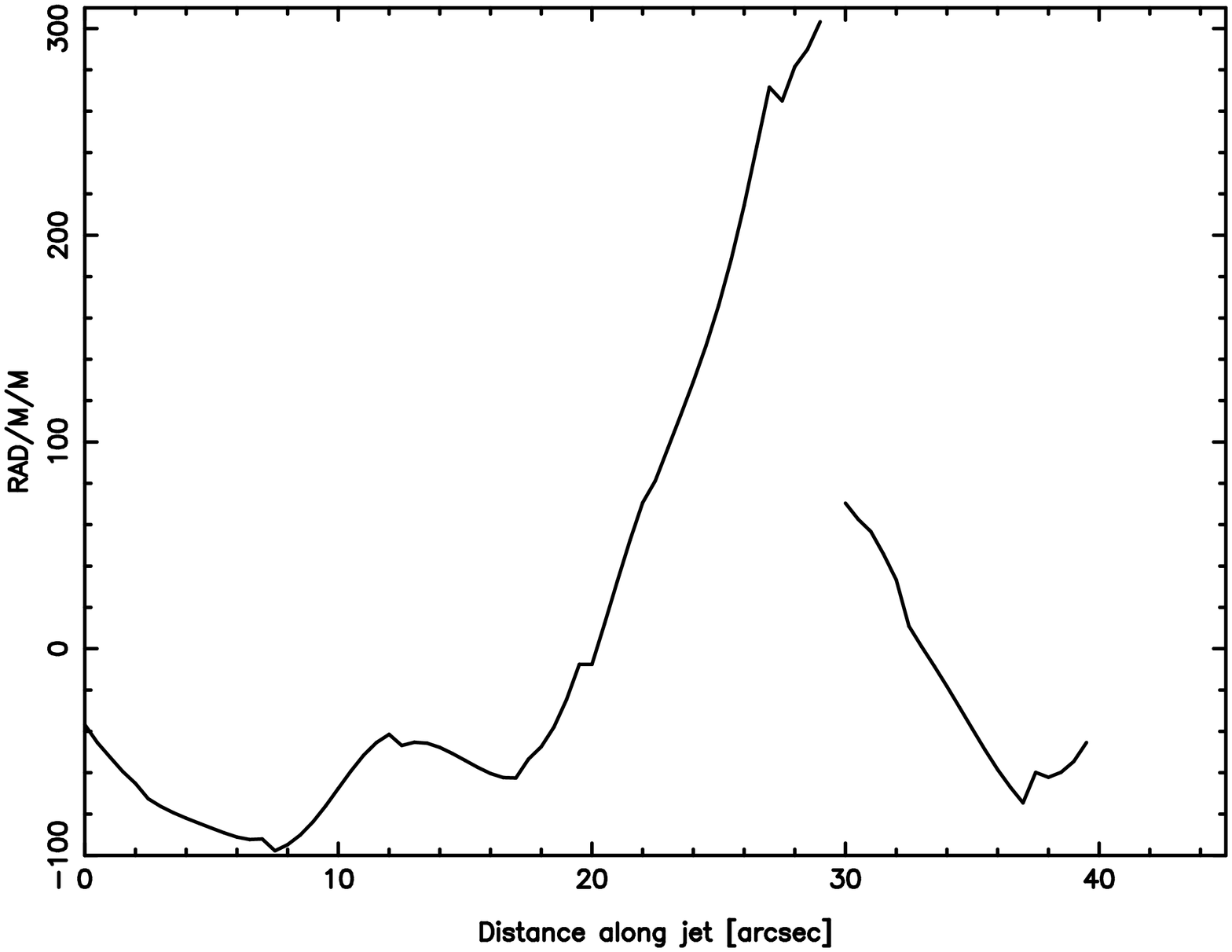}
\hfill
\includegraphics[width=0.35\textwidth,angle=270]{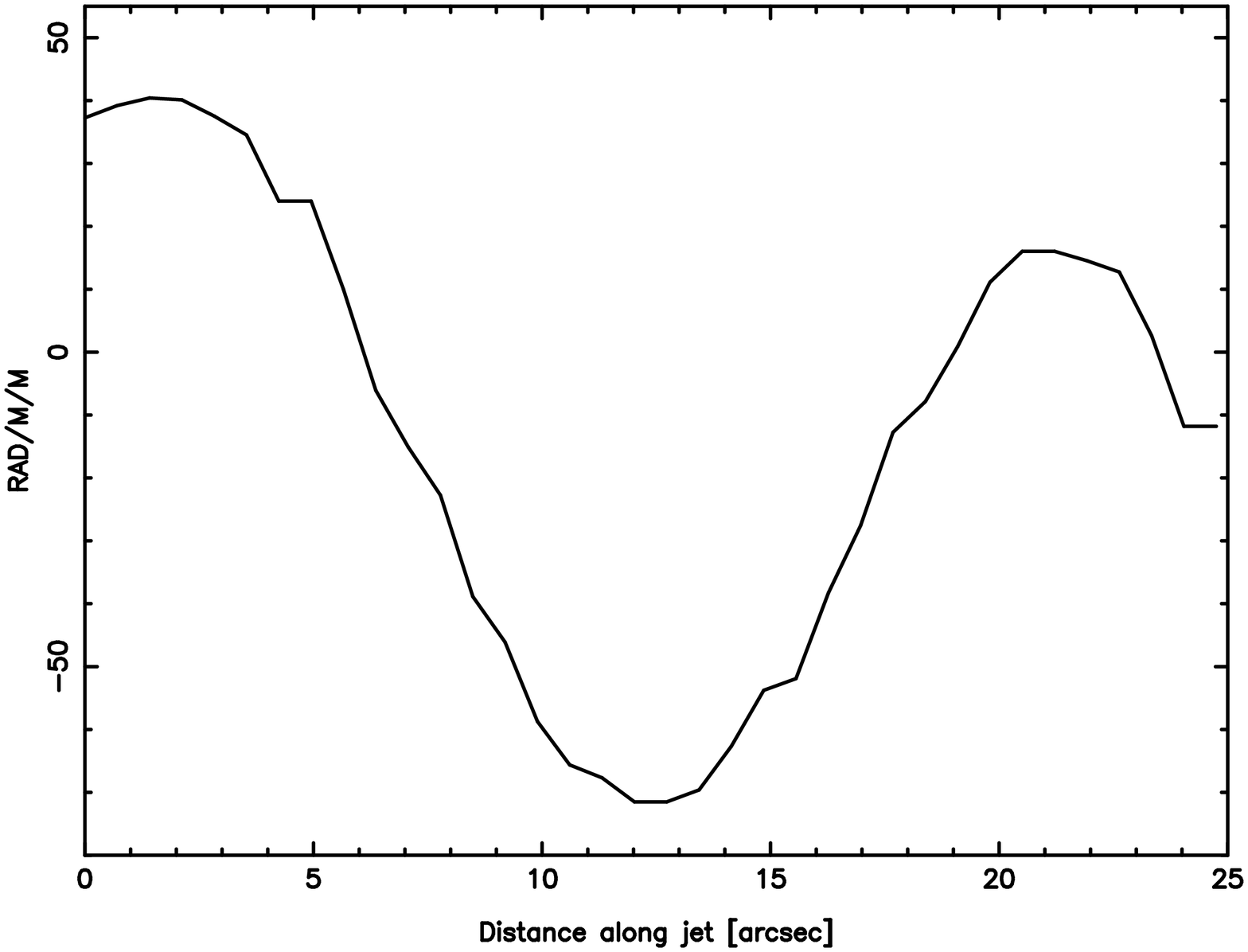}
\caption{a) Slice in the RM map of
Fig.~\ref{rm3.5cm6} at 8\arcsec\ resolution along the inner jet
(position angle $-6\degr$). The southernmost point is to the left.
The point closest to the nucleus is located at around 20 arcsec. b)
RM slice along the outer southern jet (position angle $-40.4\degr$).
The point at the southeastern jet end is to the left. The foreground
RM in our Galaxy is estimated to be $-6\pm4$~rad~m$^{-2}$.\label{rmslice}}
\end{figure*}

We present the map of Faraday rotation measures (RM) between 3.5 and
6~cm at 8\arcsec\ resolution in Figure~\ref{rm3.5cm6}. The RM at the
location of the PI maximum in the southern spiral arm is
$-60\pm20$~rad~m$^{-2}$. The RM values along the jet-like feature
are fluctuating between about $-100$ and +300~rad~m$^{-2}$
(Figure~\ref{rmslice}). The origin of Faraday rotation in NGC~7479
could be internal to the jet or in the ISM of the galaxy disk in
front of the jet, as seen by an observer. As discussed in
Section~\ref{s:origin}, the observed Faraday rotation occurs in
the foreground of the jet, either in the disk of the galaxy, in a
jet cocoon, or both.

The origin of Faraday rotation in the foreground disk is supported by
the asymmetry in
Faraday rotation which is larger by a factor of 3--5 towards the
northern jet than towards the southern jet (Fig.~\ref{rm3.5cm6}) and
also by the asymmetry in the Faraday depolarization data
(Sect.~\ref{s:dp}). The northern jet is located behind the disk of the
galaxy, while the southern jet lies in front of the disk. The jet acts
as a bright polarized background which allows us to investigate the
structure of the regular magnetic field in front of the jet with
unprecedented precision.

We took two RM slices, one along the inner jet and another along
the outer southern jet (along the two thick lines marked in
Fig.~\ref{rm3.5cm6}). The slices are shown in Figure~\ref{rmslice}.
They reveal an asymmetry in the RM values along the inner jet
(Fig.~\ref{rmslice}a). RM is negative in the south (i.e., the
regular field points away from the observer) and positive in the
north (i.e., the regular field points towards the observer). The
sharp decrease of RM values at $x=30\arcsec$ is due to the strongly
polarized outer northern jet which shows small Faraday rotation. The
field component responsible for the strong Faraday rotation of the
inner jet is probably located in the bar dust lane, whereas the
field in front of the outer jet belongs to the diffuse disk of the
galaxy and has a reversed direction with respect to the field in the
bar.

\begin{figure}[th]
\centering
\includegraphics[width=3.5in]{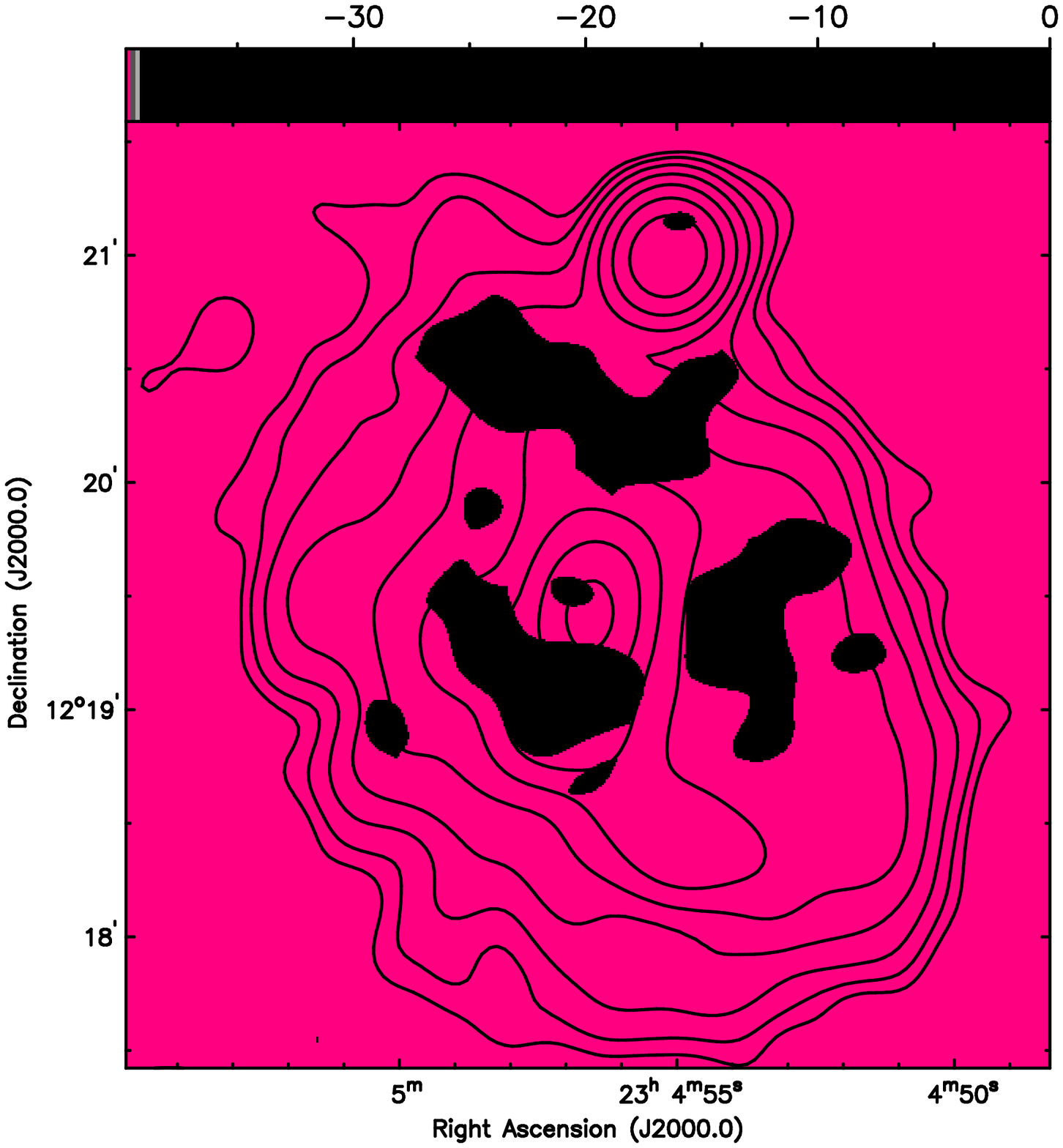}
\caption{Rotation measure map between 3.5 and 22~cm at 20\arcsec\
resolution in grayscale, overlaid by 22~cm total intensity contours.
The contour levels are at $(2, 4, 8, 16, 32, 64, 128, 256,
512)\times30~\mu$Jy~beam$^{-1}$.} \label{rm3cm22}
\end{figure}

The north--south RM asymmetry of the inner jet means that the
regular magnetic field in the foreground, probably in the bar dust
lanes, has opposite directions on the opposites sides of the
nucleus. Assuming trailing spiral arms, the signs of radial
velocities from \ion{H}{1} observations \citep{laine98} show that
the northwestern side is the near side of the disk, so that {\em the
large-scale regular field in the dust lanes points towards the
nucleus on opposite sides\/}. The same result was derived for the
barred galaxy NGC~1097 \citep{rbeck05}. Spiral galaxies also seem to
prefer inwards-directed regular fields which is a challenge for
theories of magnetic field generation in galaxies \citep{krause98}.

The 6 and 20~cm observations of \citet{ho01} with 2\farcs5
resolution show a few magnetic field vectors at the northern edge of
the inner jet ($\simeq~$10$\arcsec$ north of the nucleus) oriented
along the inner jet at both wavelengths, so that the RM seems to be
small. This is inconsistent with our results, as we measure
$RM~\simeq~$~300~rad~m$^{-2}$ in this region between 3.5 and 6~cm
(Fig.~\ref{rmslice}a). The reason for this discrepancy is probably
strong depolarization at 20~cm (see Sect.~\ref{s:dp}) and hence a
large uncertainty in the polarization angles in \citet{ho01}.

The outer southern jet (Fig.~\ref{rmslice}b) reveals variations in
RM over about 20\arcsec\ ($\simeq$~3~kpc) with sign reversals located
at $x=6\arcsec$ and 19\arcsec, indicating that the line-of-sight
component of the regular disk field in front of this region shows
{\em two reversals\/} on this scale. At the present resolution of
8\arcsec\ we cannot decide whether these reversals are smooth or
abrupt. Two more sign reversals occur between the inner and outer
southern jet (compare the first point at 0~arcsec in
Fig.~\ref{rmslice}a with the last point at 25~arcsec in
Fig.~\ref{rmslice}b) and, on a scale of about 1~kpc, also between
the inner and outer northern jet (Fig.~\ref{rmslice}a). Further
reversals may occur on scales not resolved by our observations.

We show a rotation measure map at 20\arcsec\ resolution between 3.5
and 22~cm in Figure~\ref{rm3cm22}. RM is $-30\pm3$~rad~m$^{-2}$ in
the southeast where the outer southern jet is located, about half
the RM between 3.5 and 6~cm, but of the same sign. Smaller RM
values at longer wavelengths are commonly seen in galaxies, and
indicate Faraday depolarization effects which strongly increase with
wavelength \citep{sokoloff98}.

Both sources north of the galaxy disk (at RA$_{2000}$= 23h04m55\fs2,
DEC$_{2000}=12\degr 21\arcmin 01\arcsec$, seen as an
elongated source at 12\arcsec\ resolution in Fig.~\ref{arcsec12fig})
have RM values of $-180\pm25$~rad~m$^{-2}$. The similar RM excludes
an origin internal to these sources. Their position suggests that
these are background sources, unrelated to NGC~7479, and that the RM
emerges in the disk of NGC~7479 and/or in the foreground of the
Milky Way. The negative RM means that the regular magnetic field
points away from the observer.

The polarized point-like source about 30\arcsec\ east of the nucleus
(at RA$_{2000}$= 23h04m58\fs5, DEC$_{2000}=12\degr 19\arcmin
19\arcsec$) has a rotation measure of $-20\pm20$~rad~m$^{-2}$ that
is substantially different from the northern sources. The lack of an
optical or IR counterpart indicates that this is another background
source. Assuming the same foreground RM of $\simeq~-180$~rad~m$^{-2}$
as in the north, the measured RM~$ \simeq
-20$~rad~m$^{-2}$ would require an exceptionally large internal
RM$_{\rm i}$ of $\simeq +160$~rad~m$^{-2}$. More likely, the regular
magnetic field in the disk of NGC~7479 and/or in the foreground is
reversed and points towards the observer in this region. A more
detailed investigation requires separation of the component of
Faraday rotation RM$_{fg}$ occurring in the foreground of the Milky
Way.

As an estimate of RM$_{\rm fg}$ in the direction to NGC~7479 we use
the average value of RM between 3.5 and 22~cm in the western and
northern regions, far outside the jet, which is
$-6\pm4$~rad~m$^{-2}$ (Figure~\ref{rm3cm22}). The RM sky,
interpolated from the RM grid of polarized extragalactic sources
\citep*{johnston2004}, also yields values of around 0~rad~m$^{-2}$ at
the position of NGC~7479 in Galactic coordinates (Table~1). The RM
sky shown by Han in \citet{wielebinski2005} gives small negative
values.

Knowing RM$_{\rm fg}$, we can investigate the regular magnetic field
in the disk of NGC~7479 in some more detail. The observed RM of a
background source corresponds to an average RM$_0$ through the disk of
NGC~7479 of $RM_0=(RM-RM_{\rm fg}-RM_{\rm i})/2$, while
$RM_0=RM-RM_{\rm fg}$ is valid for diffuse polarized emission from the
galaxy itself (if Faraday depolarization is small). The two northern
background sources (neglecting internal rotation RM$_{\rm i}$) give
RM$_0=-87\pm13$~rad~m$^{-2}$, which agrees (within the uncertainties)
with RM$_0=-54\pm20$~rad~m$^{-2}$ of the diffuse polarized emission in
the southern spiral arm, so that the strength and direction of the
regular field are similar in the outer northern disk and in the
southern spiral arm. However, the polarized source in the east with
RM$_0=-7\pm10$~rad~m$^{-2}$ and the multiple reversals in front of the
jet (see above) show that the large-scale magnetic field cannot have
a simple symmetry as observed in several spiral galaxies
\citep{beck05}.

Strong variations and frequent reversals of the regular field in a
galaxy disk are predicted for a mean-field dynamo with a long
timescale \citep{beck94}. Massive bars lead to non-axisymmetric gas
flows which affect dynamo action \citep{moss07}. Furthermore, shearing
flows around bars or spiral arms can generate field loops with strong
polarized emission but frequent field reversals on scales smaller than
the beamsize \citep{rbeck05}. The detection of such reversals is
generally difficult because the polarized intensities of normal barred
galaxies are low. RM reversals on $\approx 1$~kpc scale were found in
the massive spiral galaxy NGC~6946 \citep{beck07} but are hardly
significant. NGC~7479 with its bright jet is so far the most
promising laboratory for further studies at higher resolution.

\subsection{Faraday Depolarization: Turbulent Magnetic Fields}
\label{s:dp}

\begin{figure}[th]
\centering
\includegraphics[width=3.5in]{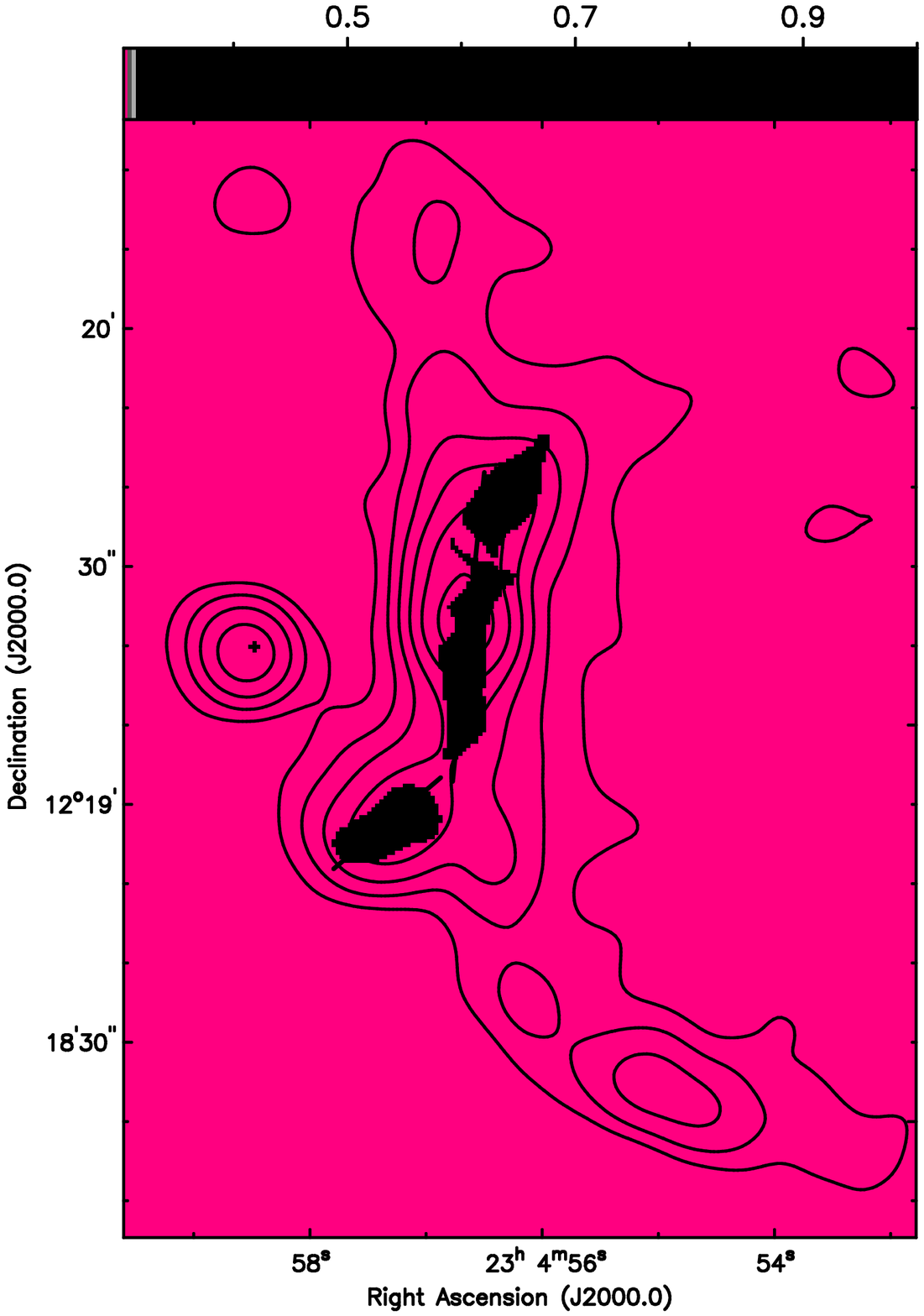}
\caption{Depolarization map between 3.5~cm and 6~cm at 8\arcsec\
resolution in grayscale, overlaid by contours of 6~cm total
intensity. The contour levels are at $(2, 4, 8, 16, 32,
64)\times50~\mu$Jy~beam$^{-1}$. The location of the slices taken
along the inner jet and the outer southern jet
(Figure~\ref{dpslice}) are indicated by thick black
lines.\label{dp35cm6}}
\end{figure}

\begin{figure}[th]
\centering
\includegraphics[width=3.5in]{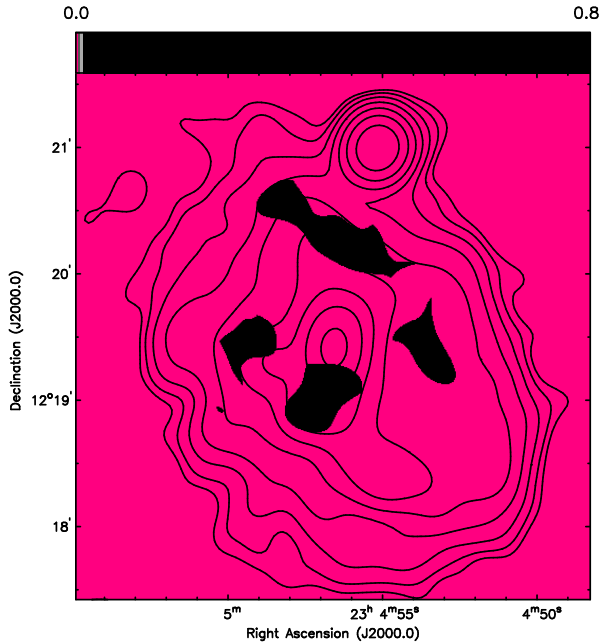}
\caption{Depolarization map between 6~cm and 22~cm at 20\arcsec\
resolution in grayscale, overlaid by contours of 22~cm total
intensity. The contour levels are at $(2, 4, 8, 16, 32, 64, 128,
256, 512)\times35~\mu$Jy~beam$^{-1}$.\label{dp6cm22}}
\end{figure}

\begin{figure}[th]
\includegraphics[width=0.35\textwidth,angle=270]{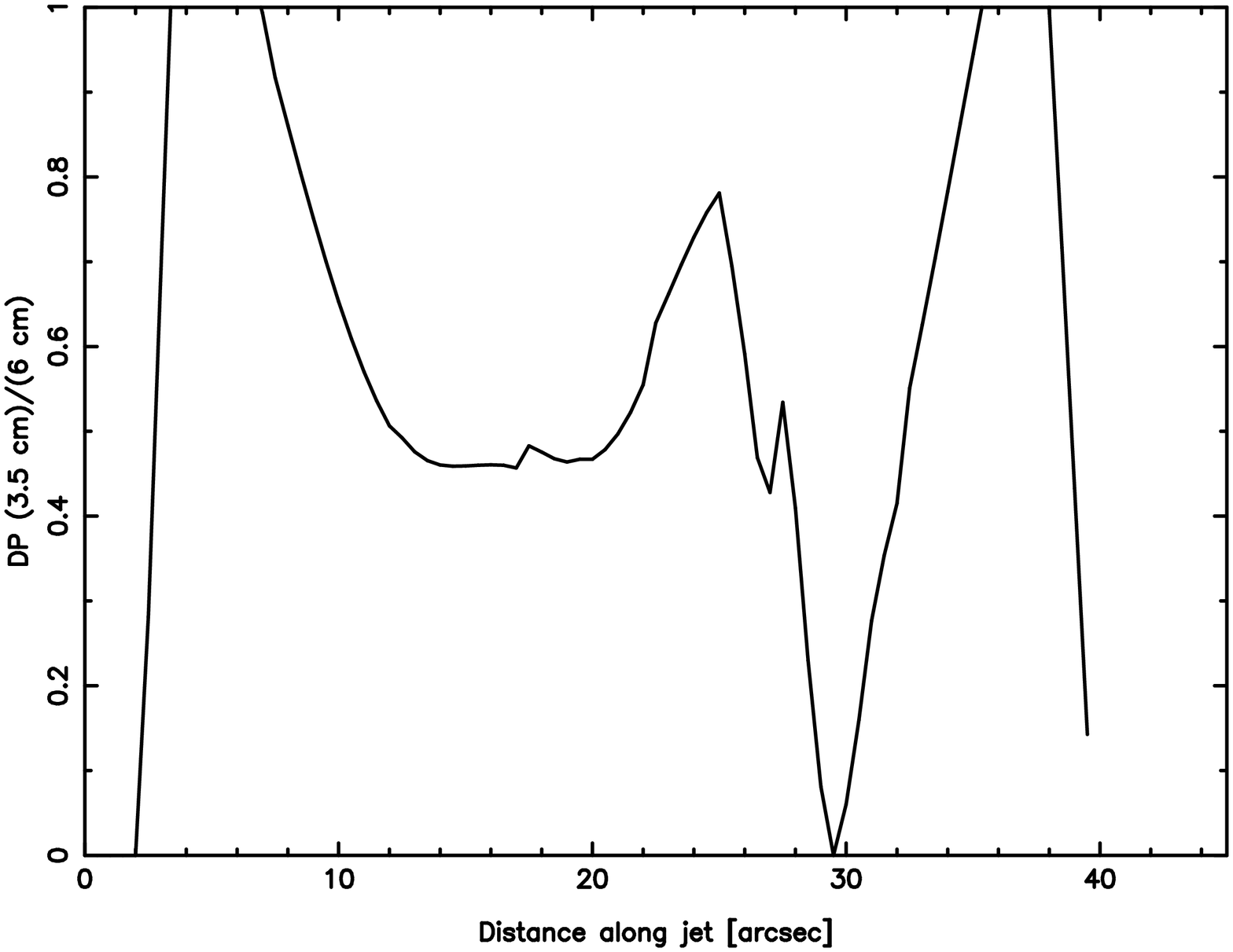}
\hfill
\includegraphics[width=0.35\textwidth,angle=270]{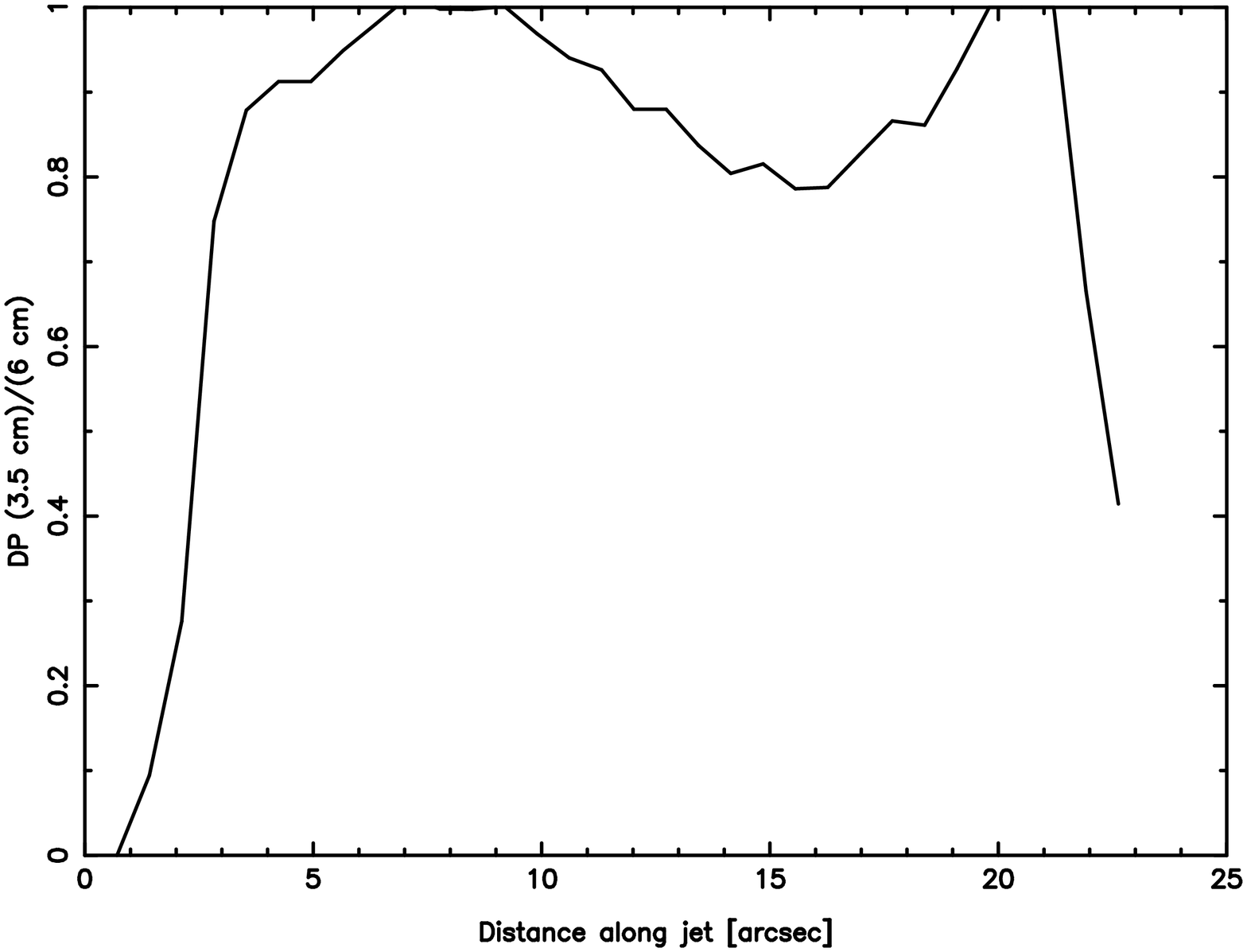}
\caption{a) Slice in the DP map of Fig.~\ref{dp35cm6} at
8\arcsec\ resolution along the inner jet (position angle
$-6\degr$). The southernmost point is to the left. b)
$DP$ slice along the outer southern jet (position angle
$-40.4\degr$). The point at the southeastern bar end is to the left.
The positions of the slices are the same as in
Figure~\ref{rmslice}.} \label{dpslice}
\end{figure}

Wavelength-dependent Faraday depolarization (DP) is a powerful tool
to detect ionized gas embedded in regular or turbulent magnetic
fields \citep{sokoloff98}. DP data are crucial for an investigation
of the origin of Faraday rotation: DP may be caused by internal
differential Faraday rotation in the regular magnetic fields within
the synchrotron-emitting medium, by internal Faraday dispersion by
turbulent fields within the emitting medium, or by external Faraday
dispersion by turbulent fields in the foreground medium. In the
first case, DP and $|RM|$ are anti-correlated, otherwise they are
independent.

Faraday depolarization is usually defined as the ratio DP of the
degrees of polarization of the synchrotron emission at two
wavelengths. This requires subtraction of the thermal emission which
is unreliable with data at only two radio frequencies. Instead, DP
in this paper was computed by DP~=~$(PI_1/PI_2) \times
(\nu_2/\nu_1)^{\alpha_{\rm n}}$ where $\alpha_{\rm n}=-1.0$ is the
synchrotron spectral index, assumed to be constant across the
galaxy. Deviations in $\alpha_n$ affect DP less severely than the
uncertainty of thermal fraction.

Figures~\ref{dp35cm6} and \ref{dp6cm22} show depolarization (DP)
values between 3.5~cm and 6~cm and between 6~cm and 22~cm,
respectively. DP along the jet (Fig.~\ref{dp35cm6}) varies between
0.6 and 1 and shows fluctuations which are not correlated or
anti-correlated with those of RM (Fig.~\ref{rmslice}). Thus, Faraday
rotation and depolarization must occur in front of the jet.

The asymmetric DP between 6 and 22~cm in the jet of NGC~7479
(Fig.~\ref{dp6cm22}) is an indication that the northern jet lies
behind the galaxy disk and the polarized emission at 18~cm and 22~cm
is depolarized by the ISM in the disk. \ion{H}{1} absorption
measurements could reveal more clearly which side of the jet-like
feature is behind the disk. Unfortunately, the observations of
\citet{laine98} revealed little neutral atomic \ion{H}{1} gas within
the bar radius of the nucleus.

Internal RM dispersion along the line of sight can be described as
$\sigma_{RM}=0.81 n_e B_r d (L f/d)^{0.5}$ \citep{sokoloff98} where
$n_e$ is the thermal electron density (in cm$^{-3}$), $B_{\rm r}$
the isotropic random field strength (in $\mu$G), $L$ the pathlength
through the thermal gas (in pc), $d$ the turbulent scale (in pc),
and $f$ the filling factor of the diffuse Faraday-rotating gas.
Depolarization occurs if the number of turbulent cells $N$ along the
line of sight is large ($N=(L f/d) \gg 1$). Then the foreground
depolarization is DP~=~e$^{-2\sigma_{\rm RM}^2 \lambda^{4}}$
\citep{sokoloff98}. Values of $n_{\rm e}=0.03$~cm$^{-3}$ outside the
bar (Sect.~\ref{s:origin}), B$_{\rm r}\simeq 7~\mu$G
(Sect.~\ref{s:strength}), and standard ISM values of
$L=500~$pc/cos~$i$, $d=50~$pc, and $f=0.5$ give $\sigma_{\rm
RM}\simeq 24$~rad~m$^{-2}$. This yields negligible depolarization at
3.5~cm ($DP\simeq 0.998$) and at 6~cm ($DP\simeq 0.98$), consistent
with the data in the outer southern jet (Figure~\ref{dpslice}b), and
strong depolarization at 22~cm ($DP\simeq 0.05$), which explains the
total depolarization of the northern jet at 18~cm and 22~cm
(Figure~\ref{arcsec20polfig}). In the outer southern jet, which is
bent above the disk as seen by an observer, assuming $L\simeq
250~$pc/cos~$i$ yields $\sigma_{\rm RM}\simeq 17$~rad~m$^{-2}$ and
$DP\simeq 0.26$ at 22~cm, leaving some polarized emission at this
wavelength. However, the condition $(L f/d) \gg 1$ is not fulfilled
here so that depolarization will be small ($DP\simeq 1$), as
observed (Figure~\ref{dp6cm22}).

In the bar, we use the same random field strength as in the southern
spiral arm of B$_{\rm r}\simeq ~17\mu$G because the radio
intensities are similar in the bar and in the southern spiral arm,
and $n_{\rm e}\simeq 0.06$~cm$^{-3}$ as derived for the bar in
Sect.~\ref{s:origin}. This yields $\sigma_{\rm RM}\simeq
120$~rad~m$^{-2}$ and hence significant depolarization ($DP\simeq
0.69$) of the inner jet at 6~cm, as observed (Fig.~\ref{dpslice}a).
In summary, our simple model of Faraday dispersion by cells of
random fields can explain the depolarization data in NGC~7479.

\citet{ho01} appear to have detected polarized emission at 20~cm at
2\farcs5 resolution at both ends of the inner jet ($\simeq
$~10\arcsec\ north and south from the nucleus), while we do not
detect any polarized emission from the inner jet at 18~cm and 22~cm
at 20\arcsec\ resolution (Fig.~\ref{arcsec20polfig}). Faraday
dispersion in the foreground disk could be smaller than in our
observations if some of the random field is resolved by the beam of
\citet{ho01}. However, the resolution of their maps of about 400~pc
at the distance of NGC~7479 is still much larger than the typical
turbulence scale of the random field. One may also consider the
possibility that the brightest ridge of polarized emission in the
jet is smaller than the random cell size in the foreground, so that
depolarization would decrease with better angular resolution. More
reliable high-resolution polarization observations around 20~cm are
needed to study the field structure of the jet and depolarization
mechanisms in more detail.

\section{DISCUSSION}

\subsection{Origin of the Faraday Rotation in Front of the Jet}
\label{s:origin}

\begin{figure*}[th]
\centering
\includegraphics[width=6.5in]{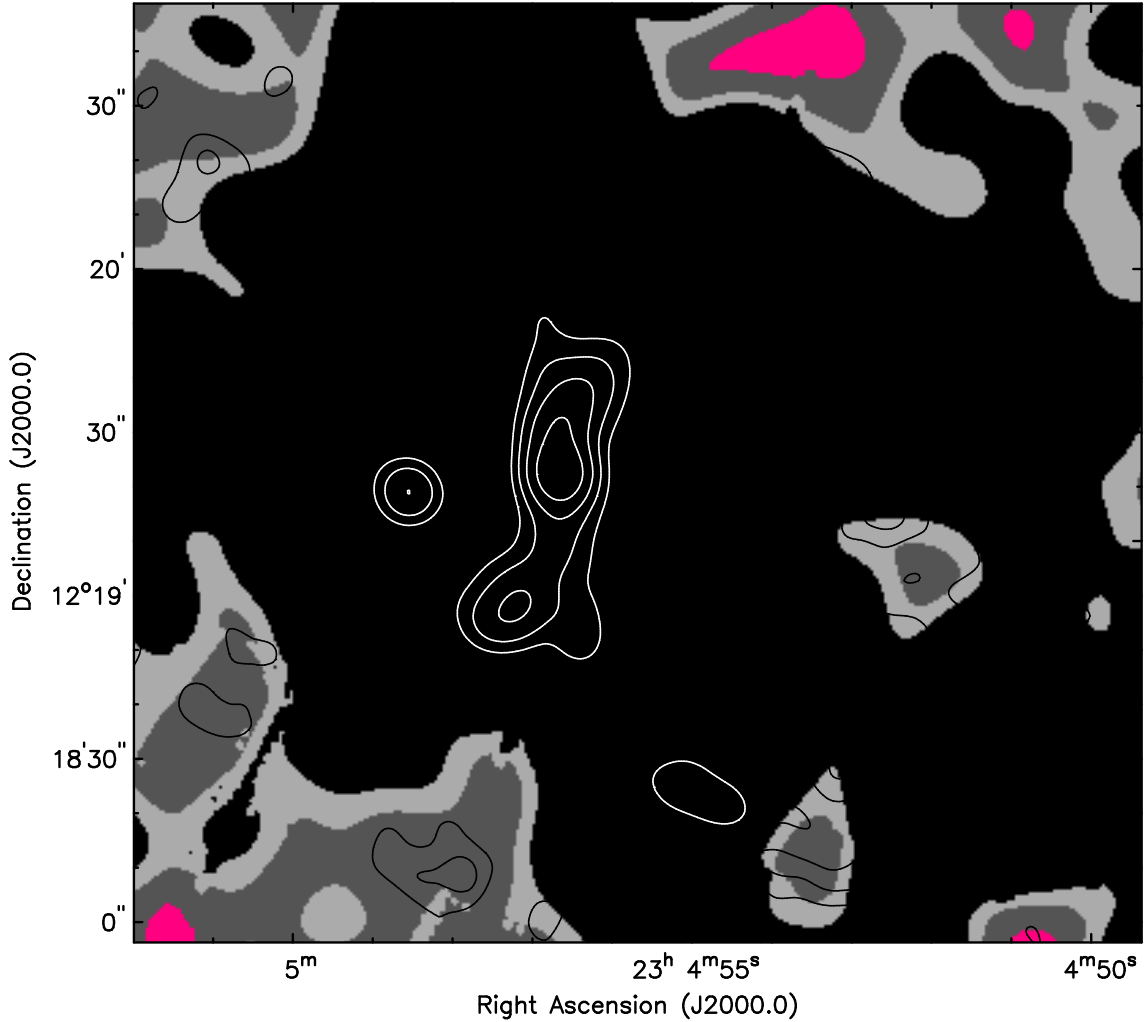}
\caption{XMM-Newton X-ray image (grayscales) in the 0.5--1.0 keV energy band of NGC~7479. The image was made by combining information from the three EPIC cameras, PN+MOS1+MOS2, and is shown in arbitrary units (M. Ehle,
private communication). The image has been smoothed with a circular Gaussian beam of 15\arcsec\ FWHM. The integration time after removal of high background periods was about 6.5 ks for PN, and 9 ks each for MOS1 and MOS2. A contour image of the 6~cm total intensity emission from Figure~\ref{arcsec8fig}b
is overlaid at 8\arcsec\ resolution. The contours are at $(2, 4, 8, 16, 32, 64, 128, 256)\times10~\mu$Jy~beam$^{-1}$.\label{xrayimage}}
\end{figure*}

The non-detection of H$\alpha$ emission from warm gas in the jet
sets an upper limit for the emission measure of about 5~pc~cm$^{-6}$
(S. Vogel, unpublished). With a pathlength of 250~pc (equal to the
jet width, see Sect.~\ref{s:maps}) this yields an upper limit of the
mean electron density of $<n_{\rm e} > ~ \le 0.14 ~
\sqrt{f}$~cm$^{-3}$ in the jet, where $f$ is the volume filling
factor of the ionized gas (for the diffuse gas $f\approx 0.5$ is
generally assumed). A two-component fit to the X-ray emission
integrated out to 30\arcsec\ radius (see Figure \ref{xrayimage} for
an XMM X-ray image of NGC~7479 with radio continuum contours
overlaid) gives a nuclear component similar to that found by
\citet{panessa06} plus a component of hot ($\approx 4\times10^6$~K)
gas with a density of $\approx 0.03/\sqrt{f}$~cm$^{-3}$ (again
assuming a pathlength of 250~pc) which is even less than the upper
limit of the warm gas. With a mean pathlength through the jet of
125~pc, $< n_{\rm e} > \le 0.1$~cm$^{-3}$ and $B_{\rm reg} \simeq
8~\mu$G give $|RM|\le 80$~rad~m$^{-2}$ from the jet which is much
smaller than observed in the inner northern jet
(Figure~\ref{rm3.5cm6}).

Our Faraday rotation data of NGC~7479 can be explained as the
effect of the magneto-ionic disk of the galaxy with a full width of
about 1~kpc if the jet is located almost in the plane of the galaxy.
Assuming that the ordered disk field with a
component in the sky plane of $8~\mu$G (Sect.~\ref{s:strength}) has
a Faraday-rotating regular component $B_{\rm reg}$ along the line of
sight of similar strength and that the effective pathlength of the
Faraday-rotating medium is 500~pc/cos~$i$ (i.e., the full thickness
of the disk, seen under an inclination of 51\degr), the observed RM
amplitude of $\simeq 300$~rad~m$^{-2}$ in front of the inner
northern jet needs a thermal electron density of $0.06$~cm$^{-3}$
which is reasonable for the bar region with a high star formation
activity. In front of the outer southern jet, RM is significantly
smaller because the jet bends out of the disk towards the observer.
Assuming the same $B_{\rm reg}$ of $8~\mu$G and an effective
pathlength of 250~pc/cos~$i$, the observation of $|RM|\simeq
70$~rad~m$^{-2}$ requires a thermal electron density of
$0.03$~cm$^{-3}$, which is typical for regions in the disks of
spiral galaxies with low star formation activity.

Our Faraday depolarization data show that the jet is bent out of the
plane, with the southern jet in front and the northern jet behind the
plane as seen by an observer (Sect.~\ref{s:dp}). However, at a mean
distance in the galaxy plane of about 6~kpc from the center, the jet
cannot be bent by more than a few degrees away from the plane,
otherwise its southern part would be located above the main part of
the thermal disk, assuming a full width of 1~kpc as is typical for
spiral galaxies, so that little Faraday rotation would be observed.

Strong interaction of the jet boundaries with the ISM gas in
the disk also explains the high synchrotron brightness of the jet
as a compression of the existing magnetic field in the disk. If a
shock exists in the boundary layer, cosmic-ray particles may also be
accelerated or re-accelerated, which would further increase the
synchrotron brightness of the jet. As the interstellar gas is also
compressed and heated in the shock, X-ray and H$\alpha$
emission is expected around the jet. More sensitive observations at
these wavelengths are needed to search for such emission.

Signs of recent perturbation by a minor merger are obvious in
NGC~7479, including the western spiral arm and the clouds of
diffuse, polarized emission between the bar and the outer spiral
arms (Figures~\ref{36cmarcsec90fig} and \ref{arcsec20polfig}).
Hence, we have to consider the possibility that the disk of ionized
gas of NGC~7479 is more extended than in non-interacting systems and
also most likely puffed up and strongly distorted
\citep[e.g.,][]{walker96}. This would allow
for somewhat larger angles between the jet and the disk plane.

Another source of Faraday rotation could be ionized gas in a cocoon
surrounding the jet and generated by instabilities in a transonic
shear layer \citep*{bick90} which would naturally explain the observed
frequent field reversals as helical field loops. However, Faraday
rotation of up to 300~rad~m$^{-2}$ and the non-detection of thermal
emission from the jet (see above) would require regular fields along
the line of sight with strengths of up to 37~$\mu$G for a pathlength
of 100~pc, much stronger than the equipartition strength of the
ordered magnetic field in the jet itself (Sect.~\ref{s:strength}),
which seems implausible. More sensitive H$\alpha$ observations are
needed to search for thermal emission from such a cocoon.

The observed RM could also emerge from a combination of the
galaxy disk and the jet cocoon in the northern jet which is
located behind the disk, while the RM in the southern jet would be solely
due to the cocoon. This scenario would allow for a large angle
between the jet and the disk. However, radio lobes and hot spots are
missing in NGC~7479. Without such a source of particle acceleration
the high radio brightness of an extraplanar jet is hard to explain.

\subsection{Field Structure in the Jets}
\label{s:field}

The north--south asymmetry in total intensity (intensity ratio of
$q\simeq 1.3$ at 3.5 and 6~cm) could be due to the viewing angle if
the northern part of the jet points towards us. However, our
depolarization data strongly suggest that the northern jet is
located behind the galaxy disk (Sect.~\ref{s:dp}) and hence points
away from us. On the other hand, the fact that the southern jet is
not brighter than the northern one is indication that the jet
velocity cannot be relativistic.

Our polarization data show a field orientation that is almost
perfectly along the jet. On the other hand, the degree of
polarization is relatively low (Sect.~\ref{s:pol}), so that a
significant part of the field is not resolved by the telescope beam
(8\arcsec\ or about 1.3~kpc in most maps shown in this paper). The
ratio of the unresolved field to the ordered field, projected into
the sky plane, is about three (Sect.~\ref{s:strength}). Such a ratio
could be explained by {\em a helical field in the jet\/}
\citep{laing81}. Radio observations with higher resolution and
higher sensitivity with the EVLA are required to determine the
profile of total and polarized intensity across the jet and hence
the field geometry and orientation with respect to the observer.

\subsection{Bending and Triggering of the Jet}
\label{s:trigger}

One of the most interesting features in our radio maps is the
reversed pitch angle of the magnetic field orientation along the
jet compared to that of the spiral arms. Such a ``leading'' spiral
feature can be generated by ram pressure of the ISM gas on the jet
near the plane with ejection velocities of about 1000~km~s$^{-1}$
\citep{sanders82}. On the other hand, the jet of NGC~4258 is presumably also
located near the plane of its host galaxy and has the same sense of
bending as the spiral arms. Ram pressure can hardly play a role for
bending the jet, hence the ejection velocities are probably larger
than 1000~km~s$^{-1}$ in NGC~7479 (see below) and also in NGC~4258
\citep{daigle01}.

Precession of the jet opposite to the rotation of the galaxy is the
most probable cause for the reversed jet bending. The precession of
the jet may be driven by the precession of the accretion disk near
the nucleus due to the Lense-Thirring effect \citep{lu90,merritt02},
by precession of the accretion disk caused by radiation
pressure \citep*{maloney96}, or by a merger event which resulted in
a binary black hole \citep*{begel80}. It has been suggested
\citep{quillen95,laine99} that NGC~7479 has recently undergone or is
undergoing a minor merger event. According to \citet{laine99} the
position of the merging nucleus is now at the bright near-infrared
and H$\alpha$ knot about 17\arcsec\ north of the nucleus.

As the jet is probably located near the plane and interacts with the
ISM gas, differential rotation of the galaxy would change the orientation
of the precessing jet to form a trailing spiral, opposite to what is
observed in NGC~7479. For a rough estimate of this effect we take the
rotation speed as 180~km~s$^{-1}$ at 7~kpc radius \citep{laine98}. To
avoid significant bending of the jet by differential rotation, the jet
propagation speed has to be larger than $\simeq$~5000~km~s$^{-1}$
which gives a travel time along the jet, and hence an age, of less than
$10^{6}$ years, consistent with the synchrotron age of the cosmic-ray
electrons (Section~\ref{s:strength}).

The anomalous 15~kpc-scale radio continuum structure that we see
emanating from the nucleus of NGC~7479 as a radio jet could be
indirectly related to the minor merger event (perturbation of the
underlying gravitational potential of the parent galaxy leading to
perhaps episodic fueling of the nucleus) which may have also created
the bar and the asymmetric spiral arm structure. The timescale of the
minor merger, several times 10$^{8}$ years, is naturally much longer
than the timescale of the jet (of order 10$^{6}$ years). Further
evidence for recent global perturbations is seen in optical images
(e.g., Fig.~\ref{arcsec2fig}) which show anomalous dust lanes
intersecting the bar at almost right angles. One such crossing point
is in the bar near the southeastern jet extension. The dust lanes come
from the western side of the bar, intersect the bar at almost right
angles, and can be followed on the eastern side of the bar for a short
distance.

It is also possible that as the jet is propagating near the plane of the
galaxy (and near the orientation of the bar), it has recently
encountered objects along its path that made it bend, so that the
jet ends moved above and below the plane. In the north, such a barrier
could be due to the merging nucleus which lies slightly north of the
point where the jet leaves the bar. In the south, it is possible that the
dust lanes trace the trail left behind by tidally stripped molecular
clouds of the disrupted companion, and they cross the bar just outside
the point where the jet bends towards southeast. CO emission is seen only
in the bar, but in the south this is close to the point where the dust
lanes cross the bar \citep{laine99b}. Thus these molecular clouds could
have blocked the passage of the jet in the bar direction, deflecting it
outside the bar. Simulations of jet deflection by a companion galaxy or
by dense clouds have been made by \citet*{wang00}, who used a jet
velocity around 4500~km~s$^{-1}$. These simulations did not attempt to
model any specific observed system, but were of general nature. To get a
deflection towards a certain direction, the impact must have taken place
off center with respect to the companion or molecular clouds. This is
difficult to verify without knowing the exact jet speed and direction,
and an accurate value for the pattern speed of the bar in which the
companion and molecular clouds are trapped. The jet is rather variable in
strength along its propagation direction, and even has sections along its
length which could be gaps. This could be due to an intermittent emission
mechanism (``blobs''), or it could be related to the
compression/deflection mechanism mentioned above.

It is unlikely that the structure that we see is an outflow or wind
from the nucleus driven by the minor merger, since we do not see any
evidence for the jet-like structure at any other than radio
continuum wavelengths (a total lack of enhanced emission outside of
the bar at other wavelengths). Furthermore, the feature is
relatively narrow and has an almost perfectly aligned magnetic
field, which would not be expected for an outflow. In fact, radio
jets are predicted to exist in galaxies with relatively weak AGN,
such as the one in NGC~7479, instead of torus outflows
\citep{moshe06}.

Future VLBI observations have the potential of revealing the
orientation of the jet near the nucleus, and thus giving further
clues to its origin \citep{scheuer96}.

\subsection{Comparison to Other Nearby Spiral Galaxies
with Large-Scale Jets}

The nearest (at around 7~Mpc) example of a spiral galaxy with a kpc-scale
radio jet is NGC~4258 \citep*{kruit72,cecil92,cecil00}. The radio jet
extends to about 14~kpc and has a spiral shape, but with the same pitch
angle as the optical spiral arms (it winds in the same sense as the
stellar spiral arms), unlike the anomalous radio continuum arms of
NGC~7479. The jet in NGC~4258 has large RM values, even larger than in
NGC~7479, possibly originating from a cocoon around the jet
\citep{krause04}. The intensities of the two jets at 3.5~cm are
comparable, several mJy per beam at a resolution of about 12\arcsec, and
the highest intensities outside the nucleus are some distance away from
the nucleus along the jet. The spectral indices are also comparable,
about $-0.7$ to $-0.8$ in both galaxies. As the jet of NGC~4258 is
narrower ($\le50$~pc) than the partly resolved jet in NGC~7479, the total
magnetic field as determined from the radio maps at 3\arcsec\ resolution
is much stronger ($\simeq 300~\mu$G). The degree of polarization in the
jet of NGC~4258 is much higher (35\%--65\%) than in NGC~7479, yielding
stronger regular fields, consistent with the higher rotation measures in
NGC~4258. Perhaps the most important difference is that the inner radio
jet in NGC~4258 has its counterpart in H$\alpha$ and X-ray emission
\citep{court61,cecil95,wilson01}, which is not seen for the
bent jet in NGC~7479, and that there is a clear interaction with the
dense ISM in the inner disk because CO line emission is seen along the
inner jet \citep{krause90,krause04,krause07}. The broader radio jet of
NGC~7479 and its weaker magnetic field strength indicate weaker interaction
with the ISM than in NGC~4258, which may explain the lack
of any signature of the jet in NGC~7479 in other spectral regimes.
The nucleus of NGC~4258 has been classified as LINER
\citep{heckman80} or a relatively weak Sy~1.9 \citep[e.g.,][]{ho97}
and therefore has a similar nuclear activity level as NGC~7479.

The edge-on disk galaxy 0313-192 at a distance of close to 300~Mpc has
huge (200~kpc) radio lobes of radio source type FR~I which extend
perpendicular to the disk \citep{ledlow01}. Only a small southern radio
jet coming from the embedded AGN \citep{keel06} has been detected, with a
length of a few kpc. The jet is weak and its total flux density is only
$729~\mu$Jy at 3.5~cm \citep{ledlow01}, corresponding to a radio power of
$\approx$~7$\times$10$^{21}$~W~Hz$^{-1}$. There is a southern jet-like
structure seen in the X-rays as well, of the same scale as the inner
radio continuum jet \citep{keel06}. No such X-ray structure has been seen
in NGC~7479. In comparison, the flux density of the southern jet of
NGC~7479 is about 3~mJy at 3.5~cm (radio power of
3.7$\times$10$^{20}$~W~Hz$^{-1}$). The main differences compared to
NGC~7479 are the existence of huge 100~kpc-scale radio lobes, the larger
jet radio power, the faster jet, and less energy loss in the inner jet of
0313-192. Due to its edge-on orientation,  it is unknown whether 0313-192
has a bar or spiral arms.

NGC~3079 is another nearly edge-on ($i=84\degr$) spiral galaxy, but at
a distance of only about 17~Mpc. It has large radio lobes
perpendicular to the disk, extending to 3~kpc above and below the
galaxy plane \citep{duric88}. However, it possesses only a small
parsec-scale nuclear radio jet, and the relation between the nuclear
jet and the large-scale radio lobes is unclear. The radio lobes are
most likely powered by a wind from the starburst nucleus, which,
however, has also LINER and Seyfert~2 classifications, and thus is a
weakly active nucleus, similar to that in NGC~7479. There is no
evidence for a strong starburst nucleus or a wind driven by a
starburst in NGC~7479. The main difference between NGC~3079 and
NGC~7479 is the lack of any kpc-scale radio jet in NGC~3079.
Therefore, the radio feature in NGC~7479 is most likely driven by the
active nucleus, whereas the radio structure in NGC~3079 is driven by
the central starburst. Other radio lobes/outflow features among nearby
edge-on spiral galaxies are seen, e.g., in the Sombrero galaxy
NGC~4594 and in NGC~4235 \citep{gallimore06}.

The nearby Circinus galaxy (distance $d\approx 4$~Mpc) has bipolar
radio lobes \citep{harnett90}, a 10~pc-scale nuclear jet-like
feature emanating from the Seyfert~2 nucleus, aligned with the radio
lobes \citep{elmo98}, and a parsec-scale nuclear outflow
\citep{green03}. Again, no kpc-scale radio jet is seen.

In summary, starburst galaxies, such as M82
\citep[e.g.,][]{seaquist91} and NGC 3079 (above) possess diffuse
radio lobes that are usually oriented perpendicular to the galaxy
disk. \citet{colbert96} found that several nearby edge-on
Seyferts galaxies also have kpc-scale diffuse radio continuum
morphologies which are not always aligned along the galaxy minor axis.
NGC~7479 does not possess a strong nuclear starburst, has a
15~kpc-scale, relatively well-collimated radio jet, and therefore differs
from both the starburst galaxies and Seyferts with diffuse kpc-scale
radio continuum structures. Thus, it represents a previously
unknown type of galaxies with a possibly interaction-triggered,
15~kpc-scale, most likely short-lived radio jet which is radio-bright
due to its location very close to the disk plane.

\section{SUMMARY}

We have examined the nature of the anomalous 15-kpc scale jet-like
feature in the nearby barred spiral galaxy NGC~7479 with radio
continuum observations at 3.5/3.6, 6, 18, and 22~cm. We have found
strong evidence to support the jet-like nature of this feature and its
origin in the nucleus of the galaxy. Most importantly, we have shown
that the large-scale radio properties of NGC~7479 differ from any other
previously mapped disk galaxy in a fundamental way. Therefore, NGC~7479
provides us a unique opportunity to study interaction-triggered,
short-lived, 15-kpc scale radio jets and large-scale magnetic fields in
disk galaxies.

Our observations suggest that the jet of NGC~7479 is located very close to
the disk plane, and thus it causes compression of the interstellar
magnetic fields and acceleration of cosmic-ray particles, resulting in an
exceptionally radio-bright jet. However, the magnetic field in the jet of
NGC~7479 is about ten times weaker than in NGC~4258, which could account
for the lack of associated extended H$\alpha$ and the weak X-ray
emission. The polarized radio emission from the jet serves as a background
against which we can measure, with the help of Faraday rotation and
depolarization, the structure of the regular magnetic field in the
foreground disk of NGC~7479 with unprecedented precision. This technique
has revealed multiple magnetic field reversals in the disk. We have
produced the most detailed study of spiral disk galaxy magnetic fields
using the jet in NGC~7479.

A more detailed summary of our main results is given below.

\begin{enumerate}

\item The total magnetic field strength is $\simeq35~\mu$G in
the inner jet and $\simeq40~\mu$G in the outer jet, which limits
the lifetime of cosmic-ray electrons to about 10$^6$ years.

\item We find that the magnetic field is closely aligned along the
radio continuum jet feature throughout its extent.

\item The degree of polarization at 3.5~cm is 6\%--8\% along the
jet, and remarkably constant, which is consistent with helical field
models for jets.

\item We can exclude the possibility that the observed Faraday rotation (RM)
between 3.5 and 6~cm and the depolarization (DP) between 6 and 22~cm
occur in the jet itself because we do not see an anti-correlation
between RM and DP (as would be expected if synchrotron emission and
Faraday rotation occur in the same volume) and no H$\alpha$ and
only weak X-ray emission is detected along the jet.

\item The observed Faraday rotation and depolarization are consistent
with magneto-ionic gas in the disk of the galaxy, between us and the
jet, with a regular field strength of about $8~\mu$G and thermal
electron densities of $\simeq 0.06$~cm$^{-3}$ near the bar and
$\simeq 0.03$~cm$^{-3}$ outside the bar. The jet cannot be bent out
of the disk scale by more than a few degrees and its true total length
is about 15 kpc. Some fraction of the Faraday effects may be generated
in a cocoon of ionized gas and regular magnetic fields around the jet.

\item The regular magnetic field in the disk of the galaxy has multiple
reversals on scales of 1--2~kpc, indicating field loops stretched by a
shearing gas flow in the bar potential of the galaxy or by gas flows
related to the merger event.

\item Faraday dispersion by cells of random field in the foreground disk
can explain the depolarization data. This model can also
account for the total depolarization of the northern jet at 22~cm if
this part of the jet is located behind the galaxy disk.

\item The ``leading'' spiral morphology of the jet, opposite to the
trailing sense of the stellar and gaseous spiral arms, may be due to precession,
driven by a precessing accretion disk near the nucleus or a binary black
hole system. Either one of these could be a consequence of a recent minor
merger event in this system. Alternatively, jet bending could be due to an
off-center collision with pieces of the companion (its nucleus and dense
molecular clouds) which happened to be in the path of the jet.

\item As the jet interacts with the ISM in the disk, the
jet must propagate at a speed of more than 5000~km~s$^{-1}$ to avoid
bending by differential rotation and hence is most likely less than
about 10$^6$ years old, and thus a very recent feature. This may explain
why such jets are not commonly seen in nearby disk galaxies.

\item The jet of NGC~7479 is the second case of a large jet located
almost in the plane of the host galaxy. Compared to NGC~4258, the jet in
NGC~7479 is broader, less polarized, hosts a weaker magnetic field, and
is bent in a sense opposite to that of the spiral arms. Both jets are
radio-bright, probably due to interaction with the dense ISM gas of the
disk, lose most of their energy within the disk, and hence cannot form
outer radio lobes.

\item More sensitive H$\alpha$ and X-ray observations are needed to
search for signs of interaction between the jet and the interstellar
medium.

\end{enumerate}

\acknowledgments

We are grateful to Vladimir Shoutenkov for help with the data reduction
and to Matthias Ehle for providing the X-ray image and for model-fitting
of the X-ray data. We thank Marita Krause for a critical reading of
the manuscript and many useful comments on an earlier version of this
paper. We thank Max Camenzind, Christian Kaiser, and Robert Laing  for
very useful discussions on jet triggering mechanisms, and Bruce
Elmegreen, Arieh K\"{o}nigl, and Tim Heckman for enlightening
discussions on the jet feature. We thank the anonymous referee for many
useful comments. The National Radio Astronomy Observatory is a facility
of the National Science Foundation operated under cooperative agreement
by Associated Universities, Inc. The Effelsberg 100-m telescope is
operated by the Max-Planck-Institut f\"ur Radioastronomie in Bonn on
behalf of the Max-Planck-Gesellschaft (MPG). This research has made use
of the NASA/IPAC Extragalactic Database (NED) which is operated by the
Jet Propulsion Laboratory, California Institute of Technology, under
contract with the National Aeronautics and Space Administration. This
work is based in part on observations made with the Spitzer Space
Telescope, which is operated by the Jet Propulsion Laboratory, California
Institute of Technology under a contract with NASA.

{\it Facilities:} \facility{VLA}, \facility{Effelsberg}, \facility{Spitzer(IRAC)}, \facility{XMM(EPIC)}.

\end{document}